\documentclass[12pt,letterpaper,a4paper]{article}
\usepackage[includeheadfoot,
marginratio={1:1,2:3},
width=551pt,
height=786pt,]{geometry}

\usepackage{amsmath}
\usepackage{amsfonts}
\usepackage{amssymb}
\usepackage{graphicx}
\usepackage{cancel}
\usepackage{empheq}
\usepackage{color}
\usepackage{hyperref}

\usepackage{graphicx}
\graphicspath{{images/}}

\numberwithin{equation}{section}
\usepackage{float}
\restylefloat{table}
\restylefloat{figure}
\usepackage[utf8]{inputenc}
\usepackage{amsmath, amsthm, amssymb, amsfonts}
\usepackage[font=small, labelfont=bf]{caption}
\usepackage{amsmath, amsthm, amssymb, amsfonts}
\usepackage{multirow}
\usepackage{graphicx}
\usepackage{float}
\usepackage[numbers,sort&compress]{natbib}
\usepackage{enumerate}
\usepackage{amsmath}
\usepackage{amsfonts}
\usepackage{amssymb}
\usepackage{graphicx}
\usepackage{cancel}
\usepackage{empheq}
\usepackage{color}
\usepackage{colortbl}
\definecolor{green2}{cmyk}{0, 1, 0.5, 0.3}
\definecolor{green3}{cmyk}{1, 0.75, 1.0, 0.0}

\definecolor{lightgreen}{cmyk}{0.2, 0, 0.2, 0.2}
\definecolor{lightgray}{cmyk}{0.1,0.2,0,0.1}
\definecolor{lightgray2}{cmyk}{0.4,0.4,0,0.8}
\definecolor{black}{cmyk}{1.0,1.0,1.0,1.0}

\usepackage{hyperref}

\restylefloat{table}
\restylefloat{figure}
\usepackage{longtable}
\usepackage{lipsum} 

\usepackage{amsfonts}
\usepackage{amssymb}
\usepackage{comment}
\usepackage{graphicx}
\usepackage{amsmath}
\usepackage{tabu}
\usepackage{dsfont}

\usepackage{amsthm}
\usepackage{slashed}
\usepackage{setspace}
\usepackage[labelformat=simple]{subcaption}

\usepackage{pgfplots}
\usepackage{tikz}
\usepackage{tikz-cd}
\usetikzlibrary{shapes.misc,shadows,fit,decorations.pathmorphing,positioning,trees,decorations.markings,decorations.pathreplacing,calc,shapes,patterns,arrows,positioning,arrows.meta}
\usepackage{xcolor}
\usepackage{pgfplots}
\usepackage{pgfplotstable}
\usepackage{xcolor}
\usepackage{makecell}
\usepackage{diagbox}
\usepackage{environ}
\usepackage[utf8]{inputenc}
\usepackage{amsmath, amsthm, amssymb, amsfonts}
\usepackage{multirow}
\usepackage{graphicx}
\usepackage{float}
\usepackage[numbers,sort&compress]{natbib}
\usepackage{enumerate}
\usepackage{enumitem}
\usepackage[capitalize,sort]{cleveref}
\usepackage{hhline}
\usepackage{mathtools}
\usepackage{longtable}
\usepackage{makecell}
\usepackage{tabularx}
\usepackage{cellspace}
\usepackage{alphalph}
\usepackage{lscape}
\usepackage{slashed}
\usepackage{setspace}
\usepackage{pifont}

\usepackage{float}

\crefname{figure}{Figure}{Figures}
\crefname{table}{Table}{Tables}

\def\be{\begin{equation}}
\def\ee{\end{equation}}
\def\bea{\begin{eqnarray}}
\def\eea{\end{eqnarray}}
\def\bes{\begin{subequations}}
	\def\ees{\end{subequations}}

\def\ov{\overline}

\def\ov{\overline}
\def\1{{\bf 1}}
\def\2{{\bf 2}}
\def\3{{\bf 3}}
\def\4{{\bf 4}}
\def\6{{\bf 6}}

\newcommand{\nn}{\nonumber}

\newcommand{\beq}{\begin{equation}}
\newcommand{\eeq}{\end{equation}}

\def\ov{\overline}

\allowdisplaybreaks[2]
\numberwithin{equation}{section}

\def\be{\begin{equation}}
\def\ee{\end{equation}}
\def\bea{\begin{eqnarray}}
\def\eea{\end{eqnarray}}
\def\bes{\begin{subequations}}
	\def\ees{\end{subequations}}

\usepackage{multicol}
\usepackage{float}
\usepackage{caption}

\usepackage{setspace}

\allowdisplaybreaks[2]
\numberwithin{equation}{section}

\usepackage{amsfonts}
\usepackage{amssymb}
\usepackage{comment}
\usepackage{graphicx}
\usepackage{amsmath}
\usepackage{tabu}
\usepackage{dsfont}
\usepackage{float}
\usepackage{amsthm}
\usepackage{slashed}
\usepackage{setspace}
\usepackage{pgfplots}
\usepackage{tikz}
\usepackage{tikz-cd}
\usetikzlibrary{shapes.misc,shadows,fit,decorations.pathmorphing,positioning,trees,decorations.markings,decorations.pathreplacing,calc,shapes,patterns,arrows,positioning,arrows.meta}
\usepackage{xcolor}
\usepackage{pgfplots}
\usepackage{pgfplotstable}
\usepackage{xcolor}
\usepackage{makecell}
\usepackage{diagbox}
\usepackage{environ}

\usepackage{amsmath, amsthm, amssymb, amsfonts}
\usepackage{amsmath, amsthm, amssymb, amsfonts}
\usepackage{multirow}
\usepackage{graphicx}
\usepackage{float}
\usepackage[numbers,sort&compress]{natbib}
\usepackage{enumerate}
\usepackage{enumitem}
\usepackage{hhline}
\usepackage{mathtools}
\usepackage{longtable}
\usepackage{alphalph}

\usepackage{amsfonts}
\usepackage{amssymb}
\usepackage{comment}
\usepackage{graphicx}
\usepackage{amsmath}
\usepackage{tabu}
\usepackage{dsfont}
\usepackage{float}
\usepackage{amsthm}
\usepackage{slashed}
\usepackage{setspace}

\usepackage{amsmath, amsthm, amssymb, amsfonts}
\usepackage[font=small, labelfont=bf]{caption}
\usepackage{amsmath, amsthm, amssymb, amsfonts}
\usepackage{multirow}
\usepackage{graphicx}
\usepackage{float}
\usepackage[numbers,sort&compress]{natbib}
\usepackage{enumerate}

\usepackage{amsmath}
\usepackage{amsfonts}
\usepackage{amssymb}
\usepackage{graphicx}
\usepackage{cancel}
\usepackage{empheq}
\usepackage{color}
\usepackage{hyperref}

\usepackage{float}
\usepackage{framed}
\restylefloat{table}
\restylefloat{figure}

\usepackage[margin=1cm]{caption}

\def\ov{\overline}

\allowdisplaybreaks[2]
\numberwithin{equation}{section}

\begin{document}
	{\hfill
		arXiv:2608.10082}

	\vspace{1.0cm}
	\begin{center}
		{\Large Tetra-quadric CY Threefold and Assisted Fibre Inflation
        }
		\vspace{0.4cm}
	\end{center}

\vspace{0.35cm}

 	\begin{center}
 		Swagata Bhattacharyya$^\ast$, George K. Leontaris$^\dagger$ and
 		Pramod Shukla$^\ast$ \footnote{Emails:~swagata.bhattacharyya@jcbose.ac.in, leonta@uoi.gr, pshukla@jcbose.ac.in}
	\end{center}
 \vspace{0.1cm}

 \begin{center}
 {$^\ast$ Department of Physical Sciences, Bose Institute,\\
 Unified Academic Campus, EN 80, Sector V, Bidhannagar, Kolkata 700091, India}\\
  \vspace{0.3cm}
{$^\dagger$ Physics Department, University of Ioannina, University Campus, \\
 Ioannina 45110, Greece}\\
  \vspace{0.3cm}

 \end{center}

\vspace{1cm}

\abstract{
In the context of type IIB superstring compactification, we demonstrate the assisted fibre inflation proposal for a four-field model realized using the orientifold of a tetra-quadric Calabi-Yau (CY) threefold. This CY threefold has an underlying permutation symmetry $S_4$ and belongs to both the list of CY threefolds, namely the Kreuzer-Skarke (KS) database as well as the Complete Intersection Calabi-Yau (CICY) database. After fixing the overall volume modulus of the CY threefold using the framework of perturbative large volume scenario, there are three K\"ahler moduli which remain flat and assist in driving fibre inflation via sub-leading corrections. In a particular benchmark model, we show that the effective inflaton shift of around $5.7$ M$_p$, as needed for driving fibre inflation, can be successfully shared by three inflaton moduli which need to be individually shifted by nearly $2.2$ M$_p$ only ! The main motivation for the work is to show that the effective large field excursions of the inflaton field is possible without the need of pushing the individual volume moduli towards a large super-Planckian excursion or close to the boundary of the K\"ahler cone which may create various subsequent challenges for the effective field theory and supergravity approximations, especially in Swiss-Cheese based models of fibre inflation.
}

\clearpage

\tableofcontents


\section{Introduction}
\label{sec_intro}
Effective Field Theories (EFTs) derived from String Theory provide a fertile ground for constructing cosmological inflationary models. In the standard quantum field theory approach, a popular and quite successful model is the slow-roll inflation~\cite{Starobinsky:1980te,Guth:1980zm,Linde:1981mu}. This occurs when a scalar field, the inflaton, rolls down a shallow region of its potential, generating the required number of efolds of accelerated expansion. During this process, quantum fluctuations of the inflaton field are stretched to superhorizon scales, seeding the primordial density perturbations that eventually give rise to the large-scale structure of the universe. After inflation ends, the inflaton field oscillates around the minimum of its potential and decays into Standard Model particles through the process of reheating, thereby initiating the hot Big Bang phase. The efficiency of reheating is typically characterized by the reheating temperature, which depends on the inflaton decay width and the equation of state during the post-inflationary epoch.

In string theory EFTs, a large number of scalar fields (collectively known as moduli) arise from the compactification of extra dimensions. These moduli parameterize the size and shape of the internal Calabi-Yau (CY) manifold, as well as the background fluxes and brane configurations. A critical challenge for constructing viable low-energy effective models is the stabilization of all string moduli. If these moduli remain undetermined, they would lead to massless or extremely light scalar fields that mediate unobserved long-range forces, alter fundamental constants, and generally render the theory phenomenologically and cosmologically inconsistent. Such a scenario would preclude the possibility of making meaningful predictions for low-energy observables. This necessitates a central task for model building known as {\it moduli stabilization} leading to two popular schemes, namely {\text KKLT} \cite{Kachru:2003aw} and the {\text LARGE Volume Scenarios (LVS)} \cite{ Balasubramanian:2005zx}.  It turns out that complex structure moduli and axio-dilaton are first stabilized by fluxes \cite{Dasgupta:1999ss,Gukov:1999ya, Taylor:1999ii,Blumenhagen:2003vr}, while the K\"ahler moduli are stabilized by using a series of perturbative and non-perturbative effects \cite{Becker:2002nn,Witten:1996bn,Green:1997di,Blumenhagen:2009qh, Blumenhagen:2008zz, Bianchi:2011qh,Bianchi:2012pn,Louis:2012nb}. We note that both, the KKLT and LVS schemes, necessarily use the non-perturbative effects, which may not be available/allowed in a given specific orientifold construction. In this regard, an alternative to moduli stabilization has been proposed using perturbative effects only \cite{Antoniadis:2018hqy,Antoniadis:2019rkh,Antoniadis:2020ryh} which include:
\begin{itemize}

\item 
BBHL's $\alpha^\prime$ correction \cite{Becker:2002nn}

\item
Log-loop correction \cite{Antoniadis:2018hqy,Antoniadis:2019rkh,Antoniadis:2020ryh,Leontaris:2022rzj,Leontaris:2025xit}

\item
A variety of string-loop corrections known as KK-type and Winding-type \cite{Berg:2004ek,vonGersdorff:2005bf, Berg:2005ja, Berg:2005yu, Cicoli:2007xp, Gao:2022uop}

\item
Higher derivative F$^4$-corrections \cite{Ciupke:2015msa}

\end{itemize}
It was shown that logarithmic string loop corrections along with the BBHL correction can help realizing an AdS minimum with exponentially large VEV for the overall volume ${\cal V}$ of the compactifying sixfold background. In fact, similar to the standard LVS, it turns out that one can have $\langle {\cal V} \rangle \propto e^{c_1/g_s^2}$ where $g_s$ is the string coupling and $c_1 \simeq {\cal O}(1)$ positive constant \cite{Leontaris:2022rzj}. For that reason it is referred as ``perturbative LVS" \cite{Antoniadis:2018hqy}. 

A major task in string cosmology is to identify a moduli potential which, preferably, exhibits a de Sitter vacuum and, upon minimization, fixes the moduli at their vacuum expectation values with positive definite masses, ensuring the stability of the vacuum. Furthermore, if the scalar potential possesses the appropriate shape, one or more moduli can play the role of the inflaton field, thereby realizing an inflationary scenario within a consistent ultraviolet-complete framework. This approach offers a unique advantage: since the moduli potential is derived from first principles in string theory, the inflationary dynamics are tightly constrained by the underlying geometry, topology, and flux choices of the compactification. Consequently, such models provide a direct link between observational cosmology and the microscopic structure of spacetime. A significant amount of efforts has been made in this direction regarding constructing (string-inspired) inflationary scenarios, e.g. see \cite{Cicoli:2023opf,McAllister:2023vgy} and references therein.

While much of the mostly early literature focuses on single-field inflationary models within the EFT framework, the canonical moduli fields in string theory typically come with exponential potentials that may naturally support the framework of assisted inflation \cite{Liddle:1998jc}. This motivates one to explore the possibility of realizing assisted inflation in string inspired models where multiple scalar fields could collectively drive the inflationary dynamics. One may hope that such a scenario could arise naturally in string theory, where a plethora of moduli fields are present. However, it is quite a challenging task as multi-field scalar potential may receive unwanted contributions from various series of corrections depending on the additional moduli, and subsequently can pose a threat to the inflationary dynamics. Therefore, a particular alignment or symmetry is needed to design a working multi-field model of inflation \cite{Leontaris:2025hly}. Nevertheless, it is indeed true that as compared to single-field inflation, assisted inflation offers several significant advantages and one should make attempts to successfully realize it within string inspired setup.

First, assisted inflation alleviates/facilitates the fine-tuning requirements inherent in single-field models. In single-field inflation with trans-Planckian field excursions, the potential must be exceptionally flat over a large field range to satisfy the slow-roll conditions and produce the observed amplitude of scalar perturbations. This typically requires delicate cancellations or special symmetries to protect the flatness of the potential against radiative corrections and higher-dimensional operators. 
In assisted inflation, by contrast, the collective dynamics of multiple fields can sustain accelerated expansion even when some or all of the individual inflaton potentials are relatively steep. This is because the overall inflationary trajectory is determined by the combined effect of all fields, and the effective potential along the multi-field path can be considerably flatter than any of the individual potentials. This feature is particularly relevant in many string constructions, where the shape of the scalar potential is dictated by first principles, such as the underlying geometry, flux quantization, and topological data of the compactification, rather than being fine-tuned to satisfy phenomenological demands of shallowness. Consequently, assisted inflation provides a more robust and theoretically well-motivated framework for realizing inflation within string theory, as it does not require the potential to be artificially flattened through parameter tuning.

Second, assisted inflation has the potential to address a major issue in the single-field inflation models in string theory, in particular, for trans-Planckian models such as fibre inflation. This is the requirement that the inflaton field takes trans-Planckian values during the inflationary dynamics. This situation is problematic for several reasons. When a field evolves over distances {\it significantly} exceeding the Planck scale, the effective field theory description breaks down, as higher-dimensional operators suppressed by the Planck mass become relevant. Such operators can significantly deform the potential, spoiling the flatness required for slow-roll inflation. Moreover, the swampland conjectures \cite{Vafa:2005ui,Ooguri:2006in}, most notably the de Sitter conjecture \cite{Obied:2018sgi,Garg:2018reu}, issues related to trans-Planckian field ranges \cite{Ooguri:2018wrx,Blumenhagen:2018nts,Grimm:2018ohb,Scalisi:2018eaz,Bedroya:2019tba} and the trans-Planckian censorship conjecture~\cite{Agmon:2022thq}, suggest that effective field theories with trans-Planckian field excursions may be incompatible with quantum gravity, residing in the swampland rather than the landscape of consistent theories. Additionally, large-field models often require special mechanisms, such as monodromy or axion alignment, to achieve super-Planckian displacements while maintaining control over the effective theory.

On the other hand, the collective action of multiple fields in assisted inflation can generate the necessary number of e-folds and match observational constraints, while each individual field takes values close to or even smaller than the Planck scale. This is because the total inflationary trajectory is shared among several fields, so the contribution of each field to the total field displacement for $n$ fields is anticipated to be reduced by some factor. In this regard, a multi-field Fibre Inflation model within the perturbative Large Volume Scenario (LVS) of type IIB string theory, has been proposed as an ``assisted inflation'' mechanism in which multiple K3-fibre moduli collectively generate sufficient e-folds \cite{Leontaris:2025hly}. Focusing on CY orientifolds with $h^{1,1}=3$, it has been shown that symmetries and sub-leading corrections (string-loop and higher-derivative effects) stabilize moduli and drive inflation entirely through perturbative effects. With these contributions, one can dispense with the need for non-perturbative terms $A\, e^{-a T}$ and exceptional divisors. Based on the observations made from the two-field assisted inflation \cite{Leontaris:2025hly}, it has been subsequently argued that in an $n$-field assisted inflation model, the individual {\it canonical} field excursions $\Delta\phi^i$ will be reduced by a factor of $\sqrt{n}$, i.e. $\Delta\phi^i\simeq \frac{\Delta\phi_{\rm total}}{\sqrt n}$ \cite{Leontaris:2025hly}.
This scaling property allows assisted fibre inflation to circumvent the issues associated with trans-Planckian field displacements while preserving the phenomenological success of the inflationary paradigm. 

In this work, we plan to extend the two-field assisted fibre inflation proposal of \cite{Leontaris:2025hly} by considering a model based on the tetra-quadric CY threefold as opposed to a toroidal cousin with volume ${\cal V} = 2 t^1 t^2 t^3$ considered earlier. This tetra-quadric CY threefold has four K\"ahler moduli along with an underlying permutation symmetry $S_4$ among the various topological ingredients. After stabilizing the overall volume modulus using perturbative LVS, the remaining three K\"ahler moduli collectively drive inflation via sub-leading string loop and higher-derivative corrections. Interestingly we find that the assistance of multiple fields is more effective in this tetra-quadric construction as compared to the toroidal-like model proposed in \cite{Leontaris:2025hly}. In fact we find that one needs only a shift of around $2.2$ M$_p$ for the three inflaton moduli in order to have an effective inflaton shift of $5.7$ M$_p$, which is even better than the anticipation of $\sqrt{n}$ factor using canonical fields in the toroidal like model !

The layout of the paper is as follows: In Section \ref{sec_setup}, we introduce the type IIB moduli fields relevant to our analysis, review the key ingredients of moduli stabilization, and describe the perturbative quantum corrections to the scalar potential. We also present a brief review of the Einstein-Friedmann field equations for the multi-field evolution during inflation. In Section \ref{sec_FI-review}, we review the Fibre inflation embedding within the perturbative Large Volume Scenario and implement it in a model based on a multi K3-fibred toroidal-like CY threefold with $h^{1,1} = 3$.  This example illustrates how the two-field approach reduces individual inflaton-field excursions of trans-Planckian displacements, a key advantage of assisted inflation.
In Section \ref{sec_tetra-quadric}, we present the relevant topological data for the tetra-quadric CY threefold and present the generic form of scalar potential arising from various corrections in this specific global CY orientifold model. Section \ref{sec_moduli-stab} presents four-field moduli stabilization with graphical demonstrations of various minima and the flat track, possibly suitable for inflation. In section \ref{sec_assisted-FI}, we extend the two-field toroidal-like assisted fibre inflation proposal of \cite{Leontaris:2025hly} to a three-field assisted fibre inflation using the tetra-quadric CY threefold. We present benchmark models to show that the individual inflaton displacements can be significantly reduced while still yielding cosmological observables in agreement with current experimental bounds. Finally in section \ref{sec_conclusions} we present our conclusions while some technical calculations are collected in the Appendix \ref{sec_appendix}. 


\section{Perturbative LVS and Model Building}
\label{sec_setup}
In the context of ${\cal N} =1$ type IIB Calabi-Yau (CY) orientifold compactifications to four dimensional supergravities, the low energy dynamics can be studied by the following quantities
\begin{itemize}
    \item the holomorphic superpotential ($W$)
    \item the real K\"ahler potential ($K$)
    \item the gauge kinetic function $(g_{\rm kin})$
\end{itemize}
The first two quantities appear through the so-called F-term contributions to the scalar potential $(V)$ which at the two-derivative level is captured by the following expressions
\be
\label{eq:V_gen}
e^{- {K}} \, V = {K}^{{\cal A} \ov {\cal B}} \, (D_{\cal A} W) \, (D_{\ov {\cal B}} \ov{W}) -3 |W|^2 \equiv V_{\rm cs} + V_{\rm k}\,,
\ee
where:
\be
\label{eq:VcsVk}
V_{\rm cs} =  K_{\rm cs}^{i \ov {j}} \, (D_i W) \, (D_{\ov {j}} \ov{W}) \qquad \text{and}\qquad V_{\rm k} =  K^{{A} \ov {B}} \, (D_{A} W) \, (D_{\ov {B}} \ov{W}) -3 |W|^2\,.
\ee
We note that the quantities $K$ and $W$ depend on the so-called chiral coordinates which are suitable for describing the underlying effective supergravity theory after performing the compactification down to four-dimensions. The chiral coordinates are defined via complexifying  various moduli with a set of RR axions; e.g. the complex-structure moduli $(U^i)$, axio-dilaton modulus ($S$) and the K\"ahler moduli $(T_\alpha)$, which play central role in typical model building, are respectively defined as $U^i = v^i - i\, u^i$, $S = c_0 + i\, s$ and $T_\alpha = c_\alpha - i\, \tau_\alpha$. Here, $s = e^{-\phi}$ is the modulus dependent on dilaton ($\phi$), $u^i$ are the saxions of the complex structure, and $\tau_\alpha$ are the four-cycle volume moduli of the Einstein frame defined as $\tau_\alpha = \partial_{t^\alpha} {\cal V} = \frac12 k_{\alpha\beta\gamma} t^\beta t^\gamma$, where ${\cal V}$ corresponds to the volume of the compactifying CY threefold. 
In addition, $c_0$ and $c_\alpha$'s are universal RR axions and RR four-form axions, respectively, while the complex structure axions are denoted by $v^i$. For our purposes, the choice of holomorphic involution is such that the indices split into two sets,  $\{i, \alpha\}$ with $i \in h^{2,1}_-({\rm CY}/{\cal O})$, and $\alpha \in h^{1,1}_+({\rm CY}/{\cal O})$,
This implies  that there are no so-called odd-moduli $G^a$ present in our analysis, and thus, $h^{1,1} = h^{1,1}_+$. We refer the interested reader to \cite{Cicoli:2021tzt} for further details on odd-moduli.

\subsection{Moduli stabilization in perturbative LVS}
The conventional moduli stabilization process in model developed within the 4D type IIB effective supergravity follows a two-step strategy. 
\begin{itemize}
\item First, one fixes the complex structure moduli $U^i$ and the axio-dilaton $S$ by the leading order flux superpotential $W_{\rm flux}$ induced by usual S-dual pair of the 3-form fluxes $(F_3, H_3)$ \cite{Gukov:1999ya}. This demands solving the following supersymmetric flatness conditions:
\bea
&& D_i W_{\rm flux} = 0 = D_{\ov {i}} \ov{W}_{\rm flux}, \qquad D_{S} W_{\rm flux} = 0 = D_{\ov {S}} \ov{W}_{\rm flux}.
\label{UStab}
\eea
After supersymmetric stabilization of axio-dilaton and the complex structure moduli, one has $\langle W_{\rm flux} \rangle = W_0$.
    
\item
At this leading order no-scale structure protects the K\"ahler moduli $T_\alpha$ which subsequently remain flat, and as a second step, they can be stabilized via including other sub-leading contributions to the scalar potential, e.g. those induced via the non-perturbative corrections in the holomorphic superpotential $W$ or the other (non-)perturbative corrections arising from the whole series of $\alpha^\prime$ and string-loop ($g_s$) corrections. 
\end{itemize}

\noindent
Unlike the standard large volume scenarios \cite{Balasubramanian:2005zx} of moduli stabilization which uses non-perturbative effects in $W$ \cite{Witten:1996bn} along with the so-called perturbative BBHL corrections in $(K)$ \cite{Becker:2002nn}, the perturbative LVS scheme uses only perturbative effects, in particular those arising from the string-loop correction of the ``log-loop"-type \cite{Antoniadis:2018hqy,Antoniadis:2018ngr} along with the BBHL correction \cite{Becker:2002nn}. There corrections are introduced through some appropriate extensions of the K\"ahler potential which schematically take the following form,
\be
\label{eq:K}
K = -\ln\left[-i\int \Omega\wedge\bar{\Omega}\right]-\ln\left[-\,i\,(S-\bar{S})\right]-2\ln{\cal Y}, 
\ee
where $\Omega$ denotes the nowhere vanishing holomorphic 3-form which depends on the complex-structure moduli and ${\cal Y}$ encodes the various K\"ahler moduli dependent pieces needed to break the no-scale structure. With the inclusion of one-loop effects of {\it log-loop type} to the K\"ahler potential on top of the BBHL corrections used in the standard LVS, one arrives at an effectively modified overall volume ${\cal V}$ which we denote as ${\cal Y}$. It takes the following explicit form,
\bea
& & {\cal Y} = {\cal Y}_0 + {\cal Y} _1, 
\eea
where ${\cal Y}_0$ denotes the overall volume modified by $\alpha^\prime$ corrections appearing at string tree-level while ${\cal Y}_1$ is induced at string one-loop level as given  below  \cite{Antoniadis:2018hqy,Antoniadis:2018ngr,Antoniadis:2019doc,Antoniadis:2019rkh,Antoniadis:2020ryh,Antoniadis:2020stf, Leontaris:2022rzj},
\bea
& & {\cal Y}_0 = {\cal V} +  \frac{\xi}{2} \, e^{-\frac{3}{2} \phi} = {\cal V} + \frac{\xi}{2}\, \left(\frac{S-\ov{S}}{2\,{\rm i}}\right)^{3/2} \,, \\
& & {\cal Y}_1 = e^{\frac{1}{2} \phi}\, f({\cal V}) = \left(\frac{S-\ov{S}}{2\,{\rm i}}\right)^{-1/2} \left(\sigma + \eta \, \ln{\cal V}\right)\,,\nonumber
\label{eq:defY}
\eea
where the various model dependent parameters $\xi$, $\sigma$ and $\eta$ are given as,
\bea
\label{eq:def-xi-eta}
& & \hskip-1cm  \xi = - \frac{\chi({\rm CY})\, \zeta[3]}{2(2\pi)^3}~, \quad \sigma  = - \frac{\chi({\rm CY})\, \zeta[2]}{2(2\pi)^3} \sigma_0, \quad \eta =  \frac{\chi({\rm CY})\, \zeta[2]}{2(2\pi)^3} \eta_0, \quad \frac{\xi}{\eta} = -\frac{\zeta[3]}{\zeta[2]}
\eea
These parameters $\xi$, $\sigma$ and $\eta$ are defined such that they do not have any string coupling ($g_s$) dependence which can be manifested by the $SL(2,\mathbb Z)$ arguments \cite{Leontaris:2022rzj}, however $\sigma$ and $\eta$ generically may depend on the complex structure moduli. In order to keep track of this possibility one typically introduces two parameters, namely $\sigma_0$ and $\eta_0$ in Eq.~(\ref{eq:def-xi-eta}) as complex-structure moduli dependence parameters while still keeping the $SL(2,\mathbb Z)$ motivated factors of the Riemann $\zeta$-functions and the Euler characteristic of the CY threefold. In fact, it is useful to introduce the following version of these parameters with some scaling using the string coupling $g_s$,
\bea
& & \hskip-1cm \hat\xi = \frac{\xi}{g_s^{3/2}}~, \qquad \hat\sigma = g_s^{1/2}\, \sigma~,\qquad \hat\eta = g_s^{1/2}\, \eta~, \qquad \frac{\hat\xi}{\hat\eta} = -\frac{\zeta[3]}{\zeta[2]\,g_s^2\, \eta_0}, \quad \frac{\hat\sigma}{\hat\eta} = -\frac{\sigma_0}{\eta_0}\,,
\eea
which subsequently results in the following leading order pieces in the scalar potential,
\bea
\label{eq:pheno-potV2}
& & V_{\alpha^\prime +{\rm log} \, g_s}^{(1)} \simeq \frac{3\, \kappa\, \hat\xi}{4\, {\cal V}^3}\, |W_0|^2 + \frac{3 \, \kappa\, (\hat\eta\ln{\cal V} - 4\hat\eta + \hat\sigma)}{2{\cal V}^3}\,|W_0|^2 \equiv V_{\rm pLVS},
\eea
where $\kappa = e^{K_{cs}}\,g_s/2$. This leading order scalar potential piece we call as $V_{\rm pLVS}$. This results in an exponentially large VEV for the overall volume determined by the following approximate relation:
\bea
\label{eq:pert-LVS}
& & \langle {\cal V} \rangle \simeq e^{\frac{13}{3}-\frac{\hat\xi}{2\, \hat\eta} -\frac{\hat\sigma}{ \hat\eta}} = e^{a/g_s^2 + b}, \qquad a = \frac{\zeta[3]}{2 \zeta[2]\eta_0} \simeq 0.365381/\eta_0, \quad b = \frac{13}{3}+\frac{\sigma_0}{\eta_0}~\cdot
\eea
For natural values $\sigma_0 = -2$ and $\eta_0 = 1$, the numerical estimate for $g_s = 0.2$ gives $\langle {\cal V} \rangle = 95594.5$ while $g_s = 0.1$ leads to $\langle {\cal V} \rangle = 7.615 \cdot 10^{16}$. Given that an exponentially large VEV of the overall volume $\cal V$ is obtained by using only the perturbative effects, this scheme is refereed as ``perturbative LVS". Moreover, similar to the standard LVS case, it corresponds to an AdS minimum.

In our current work we do not intend to add/discuss more on uplifting the Anti-de-Sitter minimum realized in the pLVS into a de-Sitter minimum, and one can simply consider the three popular classes of uplifting schemes characterized via their contributions to the scalar potential as,
\bea
\label{eq:Vuplift}
& & V_{\rm up} = \frac{{\cal C}_{\rm up}}{{\cal V}^p},
\eea
where $p = 4/3$ for anti-D3 uplifting \cite{Kachru:2003aw,Crino:2020qwk,Cicoli:2017axo,AbdusSalam:2022krp}, and $p = 2$ for D-term uplifting \cite{Burgess:2003ic,Achucarro:2006zf,Braun:2015pza} while  $p = 8/3$ for the T-brane uplifting \cite{Cicoli:2015ylx,Cicoli:2017shd}. In the effective supergravity approach, after stabilizing the overall modulus ${\cal V}$ in the perturbative LVS minimum, one can simply consider the piece as a constant shift to compensate the negative VEV of the sub-leading corrections.

\subsection{Inflation in pLVS}

There can be several types of perturbative effects that can induce interesting scalar potential contributions, e.g. useful for breaking the no-scale structure, moduli stabilisation and realizing inflationary dynamics. Apart from the leading order effects such BBHL's $(\alpha^\prime)^3$ corrections \cite{Becker:2002nn}, the perturbative `log-loop' effects of \cite{Antoniadis:2018hqy} which we have used for realizing the pLVS, there can be at least two more classes of sub-leading perturbative contributions to the scalar potential. These contributions can potentially lift all the remaining flat directions left unfixed by the pLVS and serve as possible inflaton candidates. These are some string-loop effects other than those of `log-loop' type, and the so-called higher derivative F$^4$ corrections of \cite{Ciupke:2015msa}. Let us elaborate a bit more on those:

\begin{itemize}
\item 
The so-called KK-type and Winding-type string loop effects are encoded as  corrections to the K\"ahler potential, which in the Einstein frame are written as follows \cite{Berg:2004ek, Berg:2005ja, Berg:2005yu, Berg:2007wt, Cicoli:2007xp},
\bea
\label{eq:KgsE}
& & \hskip-1cm K_{g_s}^{\rm KK} = g_s \sum_\alpha \frac{{\cal C}_\alpha^{\rm KK} \, t^\alpha_\perp}{\cal V} \,, \qquad K_{g_s}^{\rm W} =  \sum_\alpha \frac{{\cal C}_{w_\alpha}}{{\cal V}\, t^\alpha_\cap}\,.
\eea
where ${\cal C}_\alpha^{\rm KK}$ and ${\cal C}_{w_\alpha}$ are functions of the complex structure moduli and, generically, may also depend on the open-string moduli. The two-cycle volume moduli $t^\alpha_{\perp}$ denote the transverse distance among the various stacks of the non-intersecting $D7$-brane and $O7$-planes, whilst  $t^\alpha_{\cap}$ denotes the volume of the curve sitting at the intersection loci of the various non-trivially intersecting stacks of $D7$-branes such that the intersecting curve is non-contractible. These effects induce correction to the scalar potential which take the following form,
\bea
& & V_{g_s}^{\rm KK} = \kappa \, g_s^2 \, \frac{|W_0|^2}{4\,{\cal V}^4} \sum_{\alpha,\beta} {\cal C}_\alpha^{\rm KK} {\cal C}_\beta^{\rm KK} \left(2\,t^\alpha t^\beta - 4\, {\cal V} \,k^{\alpha\beta}\right), \\
& & V_{g_s}^{\rm W} = -2 \kappa \frac{|W_0|^2}{{\cal V}^3} \, \sum_\alpha \frac{{\cal C}_{w_\alpha}}{t^\alpha_\cap}, \nonumber
\eea
Such effects can generically appear at ${\cal O}({\cal V}^{-10/3})$ in the large volume expansion. In fact there can be additional loop corrections motivated by the field theoretic computations \cite{vonGersdorff:2005bf,Gao:2022uop}, however we do not include those corrections in the current analysis; see \cite{Gao:2026mvc} for an interesting work in this regard.

\item
Finally, the higher derivative F$^4$-corrections cannot be captured by the conventional two-derivative corrections through the K\"ahler potential and the superpotential. These corrections take the following simple form \cite{Ciupke:2015msa},
\bea
& & V_{{\rm F}^4} = -  \frac{\lambda\,\kappa^2\,|W_0|^4}{g_s^{3/2} {\cal V}^4} \Pi_\alpha \, t^\alpha\,, \qquad \Pi_\alpha = \int_{\rm CY} c_2({\rm CY}) \wedge \hat{D}_\alpha\,,
\eea
where $|\lambda|\simeq 10^{-4}$and is a model dependent parameter with some combinatorial factor \cite{Grimm:2017pid, Cicoli:2023njy}, and $\Pi_\alpha$ is the second Chern-number corresponding to the CY divisor $D_\alpha$. We note that such corrections naively appear at ${\cal O}({\cal V}^{-11/3})$ in the large volume expansion.

\end{itemize}

\noindent
Thus let us emphasize that using simply the Gukov-Vafa-Witten's (GVW) flux superpotential $W_0$ for stabilizing the complex-structure moduli and the axio-dilaton at their respective supersymmetric minimum, the perturbative scalar potential contributions relevant for our purpose can be collected in the following form
\bea
\label{eq:masterVgen}
& & \hskip-0.75cm V_{\rm tot} = V_{\rm up} + V_{\rm pLVS} + V_{\rm inf} \dots,
\eea
where 
\bea
& & V_{\rm inf} = V_{g_s}^{\rm KK} + V_{g_s}^{\rm W} +V_{{\rm F}^4}. 
\eea
Here we consider $V_{\rm up}$ and $V_{\rm pLVS}$ pieces to depend on the overall volume modulus ${\cal V}$ only, while $V_{\rm inf}$ can generically depend on all the moduli. After seeking a mass-hierarchy between the modulus ${\cal V}$ and the rest of the moduli, one can explore the possibility of realizing inflationary models in perturbative LVS framework.

\subsection{On realizing multi-field inflation in pLVS}
To analyze the multi-field evolution of inflationary dynamics, we adopt the e-folding number $N$ 
as the time coordinate, defined by  $dN = H dt$. In this framework, the Einstein-Friedmann equations governing the scalar fields $\Phi^a$ take the following form:

\bea
\label{EOM}
\frac{d^2 \Phi^a}{dN^2}+{\Gamma^a}_{bc} \frac{d\Phi^b}{dN} \frac{d\Phi^c}{dN}+\left(3+ \frac{1}{H} \frac{dH}{dN}\right) \frac{d\Phi^a}{dN}+ \frac{{\cal G}^{ab} \partial_b V}{H^2}=0,
\eea
where the Hubble function is given as
\bea
\label{constran}
H^2= \frac{1}{3}\left(V(\Phi^a)+ \frac{1}{2}H^2 \, {\cal G}_{ab} \frac{d\Phi^a}{dN} \frac{d\Phi^b}{dN} \right).
\eea
Here, the generic (non-flat) field space metric is represented as ${\cal G}_{ab}$ while ${\Gamma^a}_{bc}$'s denote the corresponding Christoffel connections, and the derivatives of the scalar potential are defined as: $\partial_a V = \frac{\partial V}{\partial \Phi^a}$. Furthermore, the field space metric ${\cal G}_{ab}$ for the real moduli is obtained from the K\"ahler moduli space metric $K_{T_\alpha\ov{T}_\beta}$ using the following relation
\bea
\label{eq:fieldspace-metric}
& & K_{T_\alpha\ov{T}_\beta} (\partial_\mu T_\alpha) (\partial^\mu \ov{T}_\beta) = \frac{1}{2} {\cal G}_{ab} (\partial_\mu \Phi^a) \,(\partial^\mu\Phi^b),
\eea
where $\{\Phi^a\}$ forms the new basis of the real fields.
However, since the scalar potential in the present construction does not depend on the axionic moduli, all $C_4$ axions  remain flat within the perturbative LVS framework. We therefore restrict the set $ \{\Phi^a\}$ to include only the volume moduli.
In addition, since the overall volume ${\cal V}$ is stabilized in the leading order in perturbative LVS and the conversion from two-cycle volume $t^\alpha$ to four-cycle volume $\tau_\alpha$ does not necessarily lead to a simple enough form for generic CY threefolds, we will work on the basis of real fields $\Phi^a = \{{\cal V}, t^2, .., t^n\}$.
This basis includes the overall volume ${\cal V}$, making it well-suited for preserving the mass hierarchy between the modulus ${\cal V}$ and the remaining orthogonal moduli  $t^\alpha$.

To study inflationary dynamics, we must examine the sufficient conditions for slow-roll inflation, which are determined by a set of slow-roll parameters. We define two such parameters, $\epsilon_H$ and $\eta_H$,  as follows:
\bea
\label{eq:epH-etaH}
& & \epsilon_H \equiv - \frac{\dot{H}}{H^2} = - \frac{1}{H} \frac{dH}{dN},\qquad \eta_H \equiv \frac{\dot{\epsilon_H}}{\epsilon_H\, H} = \frac{1}{\epsilon_H}\frac{d\epsilon_H}{dN}.
\eea
Moreover, we find the following useful relation, 
\bea
\label{third}
& & \epsilon_H = 3 - \frac{V}{H^2} = \frac{1}{2} \, {\cal G}_{ab} \frac{d\Phi^a}{dN} \frac{d\Phi^b}{dN} > 0.
\eea
We emphasize that $\epsilon_H \ll 1$ is a necessary condition for inflationary dynamics, alongside $\eta_H \ll 1$, to ensure the slow-roll regime produces a sufficient number of $e$-folds. These generic slow-roll parameters $(\epsilon_H, \eta_H)$ can be expressed in terms of potential-dependent parameters $(\epsilon_V, \eta_V)$ as
\bea
\label{eq:epV-etaV}
& & \epsilon_H \simeq \epsilon_V, \qquad \eta_H \simeq 4 \epsilon_V - 2 \eta_V~.
\eea
where $\epsilon_V$ and $\eta_V$ are defined via the field-space metric $\mathcal{G}^{ab}$ and the Hessian structure of the scalar potential $V$:
\bea
\epsilon_V &=& \frac{\mathcal{G}^{ab} (\partial_a V) (\partial_b V)}{2 V^2},  \\
N^a_b &=& \frac{\mathcal{G}^{ac} \left(\partial_c \partial_b V - \Gamma^d_{cb} \partial_d V\right)}{V}.\nonumber
\eea
Here, $\eta_V$ corresponds to the most negative eigenvalue of $N^a_b$. These parameters are critical for determining the cosmological observables that characterize the inflationary model.

Given that one usually starts with a scalar potential which is a function of the moduli fields, it turns out that the following form of the field equations (\ref{EOM}) is useful,
\bea
\label{eq:EOM2}
& & \frac{d^2\Phi^a}{dN^2}+{\Gamma^a}_{bc} \frac{d\Phi^b}{dN} \frac{d\Phi^c}{dN}+\left(3- \frac{1}{2} \, {\cal G}_{ab} \frac{d\Phi^a}{dN} \frac{d\Phi^b}{dN} \right) \left(\frac{d\Phi^a}{dN}+ \frac{{\cal G}^{ab} \partial_b V}{V} \right)=0~,
\eea
where the scalar potential $V$, the metric ${\cal G}_{ab}$ as well as its inverse ${\cal G}^{ab}$ are explicitly known functions of the fields $\Phi^a$.

Now, our primary objective is to solve the second-order differential equation (\ref{EOM2}) to determine the evolutionary trajectories of various fields under different initial conditions.  For instance, we can impose the initial conditions 
\bea
& & \Phi^a(0)=\Phi^a_0 \qquad  {\rm and} \qquad \frac{d\Phi^a}{dN}|_{N=0}=0~,
\eea
and numerically evolve the system until the end of inflation.

By numerically solving the second-order differential equations (\ref{eq:EOM2}), we obtain the field trajectories $\Phi^a(N)$ as functions of the number of $e$-folds. These solutions allow us to track the evolution of the scalar potential and the slow-roll parameters. From these, we can derive the cosmological observables, in particular  the scalar power spectrum $P_s$, the spectral index $n_s$, its running $\alpha_s$, and the tensor-to-scalar ratio $r$, all expressed in terms of the $e$-folding evolution.
These observables are defined as
\bea
\label{eq:cosmo-observables}
& & P_s(N) = \frac{V(N)}{24\pi^2\, \epsilon_H(N)}, \quad n_s(N) = 1 + \frac{1}{P_s(N)} \frac{d}{dN} \,P_s(N), \\
& & \alpha_s(N) = \frac{d}{dN} n_s(N), \qquad r(N) = 16 \, \epsilon_H(N). \nonumber
\eea
All the cosmological observables are evaluated at the horizon exit $\Phi^a = \Phi^{a\ast}$ with suitable initial conditions such that one typically has $N(\Phi^{a\ast}) \gtrsim 60$ as an estimate for the sufficient $e$-folds needed to ensure a successful inflationary scenario. 
However, the total number of 
$e$-folds $(N)$ depends on multiple factors, including post-inflationary dynamics, and can be expressed as a sum of several distinct contributions \cite{Liddle:2003as,Cicoli:2017axo}:
\bea
\label{eq:cosmo-observables1}
& & \hskip-0.5cm N = \int H \, dt \simeq 57 + \frac{1}{4} \ln(r_\ast V_\ast) - \frac{1}{3}\ln\left(\frac{10V_{\rm end}}{m_{\inf}^{3/2}}\right),
\eea
where $V_{\rm end}$ corresponds to the value of the potential at the end of inflation determined by $\epsilon_H = 1$ and $m_{\rm inf}$ is the inflaton mass. From the single-field analysis, one usually has $N \simeq 50$ for Fibre inflation \cite{Cicoli:2017axo,Bhattacharya:2020gnk,Cicoli:2020bao}. Following the constraints from the Planck 2018 data, one typically needs the scalar perturbation amplitude to satisfy $P_s \simeq 2.1 \times 10^{-9}$ while the primordial scalar tilt is $n_s=0.9651\pm 0.0044$ and its running turns out to be $\alpha_s = - 0.0041 \pm 0.0067$ \cite{Planck:2018jri,Planck:2018vyg}. In addition, the Atacama Cosmology Telescope (ACT) data gives $n_s= 0.9666 \pm  0.0077$, in agreement with Planck while a new constraint based on a combination of the Planck and ACT data (P-ACT) gives $ n_s = 0.9709 \pm  0.0038$. Furthermore, a combination of Planck, ACT, and DESI (P-ACT-LB) gives $n_s = 0.9743 \pm  0.0034$ and $\alpha_s = 0.0062 \pm 0.0052$ \cite{ACT:2025tim,ACT:2025fju,DESI:2024mwx, Frolovsky:2025iao}.


\section{Reviewing Fibre Inflation in Perturbative LVS}
\label{sec_FI-review}
Inflationary models realized within the framework of perturbative LVS have been studied with a special CY example which appears to resemble with the toroidal orbifold ${\mathbb T}^6/({\mathbb Z}_2 \times {\mathbb Z}_2)$ in terms of various topological properties. Before presenting a brief review of the (single- and two-field) fibre inflation models realized in pLVS, we present the relevant topological data for this K3-fibred CY threefold.

\subsection{A concrete K3-fibred CY threefold as a toroidal cousin}
This is actually a multi K3-fibred CY$_3$ with $h^{1,1} = 3$ corresponding to the polytope Id: 249 in the CY database of \cite{Altman:2014bfa} which has been studied earlier in \cite{Gao:2013pra, Leontaris:2022rzj, Bera:2024zsk}. It is described by the following toric data:
\begin{center}
\begin{tabular}{|c|ccccccc|}
\hline
\cellcolor[gray]{0.9}Hyp &\cellcolor[gray]{0.9} $x_1$  &\cellcolor[gray]{0.9} $x_2$  &\cellcolor[gray]{0.9} $x_3$  &\cellcolor[gray]{0.9} $x_4$  &\cellcolor[gray]{0.9} $x_5$ & \cellcolor[gray]{0.9}$x_6$  &\cellcolor[gray]{0.9} $x_7$  \\
\hline
\cellcolor[gray]{0.9}4 & 0 & 0 & 1 & 1 & 0  & 0 & 2 \\
\cellcolor[gray]{0.9}4 & 0 & 1 & 0 & 0 & 1  & 0 & 2 \\
\cellcolor[gray]{0.9}4 & 1 & 0 & 0 & 0 & 0  & 1 & 2 \\
\hline
& K3  & K3 & K3 &  K3 & K3 & K3 & SD  \\
\hline
\end{tabular}
\end{center}
The Hodge numbers are $(h^{2,1}, h^{1,1}) = (115, 3)$, the Euler number is $\chi=-224$ while the Stanley-Reisner ideal is ${\rm SR} =  \{x_1 x_6, \, x_2 x_5, \, x_3 x_4 x_7 \}$. The analysis of the divisor topologies using {\it cohomCalg} \cite{Blumenhagen:2010pv, Blumenhagen:2011xn} shows that the first six toric divisors are K3 surfaces while the seventh one is described by Hodge numbers $\{h^{0,0} = 1, h^{1,0} = 0, h^{2,0} = 27, h^{1,1} = 184\}$. Considering the divisor basis $\{D_1, D_2, D_3\}$ and the K\"ahler form $J = t^1 \hat{D}_1+t^2 \hat{D}_2+t^3 \hat{D}_3$, the second Chern class $c_2({\rm CY})$ is given as 
\be
c_2({\rm CY}) = 5 \hat{D}_3^2+12 \hat{D}_1 \hat{D}_2 + 12 \hat{D}_2 \hat{D}_3+12 \hat{D}_1 \hat{D}_3,
\ee
while $k_{123} = 2$ is the only non-zero triple intersection leading to 
\be
{\cal V} = 2 \, t^1\, t^2\, t^3 = \frac{1}{\sqrt{2}}\,\sqrt{\tau_1 \, \tau_2\, \tau_3},
\ee
where $\tau_1 = 2 t^2 t^3, \tau_2 = 2 t^1 t^3, \tau_3 = 2 t^1 t^2$ and ${\cal V} = t^1 \tau_1 = t^2 \tau_2  = t^3 \tau_3$ as in toroidal case. The K\"ahler cone conditions are:
\bea
\label{eq:KC-torus}
& & \hskip-1.5cm \text{KCC:} \qquad  t^1 > 0, \quad t^2 > 0, \quad t^3 > 0.
\eea

\noindent
{\bf Brane setting:} For a given holomorphic involution, one needs to introduce D3/D7-branes and fluxes in order to cancel all the D3/D7 tadpoles. In fact, one can nullify D7-tadpoles by introducing stacks of $N_a$ D7-branes wrapped around suitable divisors (say $D_a$) and their images ($D_a^\prime$). However, the presence of D7/O7 also contributes to the D3-tadpoles, which receive contributions from  $H_3/F_3$ fluxes, D3-branes and O3-planes. The resulting D3/D7 tadpole cancelation conditions are given as \cite{Denef:2004cf,Denef:2004ze,Blumenhagen:2008zz}:
\bea
\label{eq:D3D7tadpole}
& & \hskip-0.5cm {\bf \rm D7:} \,\, \sum_a\, N_a \left([D_a] + [D_a^\prime] \right) = 8\, [{\rm O7}],\\
& & \hskip-0.5cm {\bf \rm D3:} \,\, N_3 = \frac{N_{\rm O3}}{4} + \frac{\chi({\rm O7})}{12} + \sum_a\, \frac{N_a \left(\chi(D_a) + \chi(D_a^\prime) \right) }{48},\nonumber
\eea
where $N_3 \equiv N_{\rm D3} + \frac{N_{\rm flux}}{2} + N_{\rm gauge}$ such that $N_{\rm D3}$ is the net number of D3-brane, $N_{\rm flux} = (2\pi)^{-4} (\alpha^\prime)^{-2}\int_X H_3 \wedge F_3$ is the contribution from background fluxes and $N_{\rm gauge} = -\sum_a (8 \pi)^{-2} \int_{D_a}\, {\rm tr}\, {\cal F}_a^2$ is due to D7 worldvolume fluxes. Considering the involution $x_7 \to - x_7$ results in fixed point set $\{O7 = D_7\}$ without any $O3$-planes, and subsequently one can consider 3 stacks of $D7$-branes wrapping each of the three divisors $\{D_1, D_2, D_3\}$, and the D3/D7-brane tadpoles are nullified via $8\, [O_7] = 8 \left([D_1] + [D_1^\prime] \right) + 8 \left([D_2] + [D_2^\prime] \right)+ 8 \left([D_3] + [D_3^\prime] \right)$ resulting in $N_3 = 44$. This number appears to be large enough to accommodate the required flux
choice and yield a suitable value of the $W_0$ parameter.


\noindent
{\bf Sub-leading corrections to scalar potential :} Given that there are no rigid divisors present, a priory this setup will not receive non-perturbative superpotential contributions \cite{Witten:1996bn} from instanton or gaugino condensation. The divisor intersection analysis shows that all the three $D7$-brane stacks intersect at ${\mathbb T}^2$ while each of those intersect the $O7$-plane on a curve ${\cal H}_9$ defined by $h^{0,0} = 1$ and $h^{1,0} = 9$. Further, there are no non-intersecting $D7/O7$ stacks and no $O3$-planes, and therefore this model does not induce the KK-type string-loop corrections to the K\"ahler potential \cite{Berg:2004sj,vonGersdorff:2005bf,Berg:2005ja,Berg:2007wt,Cicoli:2007xp,Gao:2022uop}. However, as the $O7/D7$ stacks intersect on non-shrinkable 2-cycles with size $t_\cap^i$, one will have string-loop effects of the winding-type \cite{Berg:2004sj,vonGersdorff:2005bf,Berg:2005ja,Berg:2007wt,Cicoli:2007xp,Gao:2022uop}: $V_{g_s}^{\rm W} = -\frac{\kappa |W_0|^2}{{\cal V}^3}\sum_i\frac{{\cal C}_{wi}}{t^i_\cap}$ where $C_{wi}$ are CS-moduli dependent coefficients. This CY has several properties similar to a toroidal case, however basis divisors being $K3$ implies that their second Chern numbers are non-zero, $\Pi(D_\alpha) = 24$,  unlike ${\mathbb T}^4$ (see \cite{Carta:2022web} for CY threefolds with a ${\mathbb T}^4$ divisor), leading to the higher derivative F$^4$ corrections to the scalar potential $V_{{\rm F}^4} \propto \Pi_\alpha t^\alpha$ \cite{Ciupke:2015msa}. Summarising all the contributions, we have \cite{Bera:2024zsk}:
\bea
\label{eq:Vfinal-simp}
& & \hskip-0.75cm V = V_{\rm up} +  \frac{{\cal C}_1}{{\cal V}^3} \left(\hat\xi + 2\,\hat\eta \, \ln{\cal V} - 8\,\hat\eta + 2\,\hat\sigma \right) \\
& & \hskip-0.35cm  - \frac{\kappa\,|W_0|^2}{{\cal V}^3} \left(\frac{{\cal C}_{w_1}}{t^1} + \frac{{\cal C}_{w_2}}{t^2} +\frac{{\cal C}_{w_3}}{t^3} +\frac{{\cal C}_{w_4}}{2(t^1+t^2)} +\frac{{\cal C}_{w_5}}{2(t^2+t^3)} +\frac{{\cal C}_{w_6}}{2(t^3+t^1)} \right) \nonumber\\
& &  \hskip-0.35cm  \, +  \frac{{\cal C}_3}{{\cal V}^4}\,\left(t^1 + t^2 + t^3 \right) + \cdots,\nonumber
\eea
where the various coefficients ${\cal C}_i$'s are given as,
\bea
\label{eq:calCis}
& & \hskip-1cm {\cal C}_1 = \frac{3\,\kappa\, |W_0|^2}{4} = \frac34 {\cal C}_2, \quad {\cal C}_3 = - \frac{24\, \lambda\,\kappa^2\, |W_0|^4}{g_s^{3/2}}, \quad |\lambda| =  \, {\cal O}(10^{-4}), \quad \kappa = \frac{g_s}{2}e^{K_{cs}}\cdot
\eea
As ${\cal V} = 2 \,t^1\, t^2\, t^3$, the scalar potential (\ref{eq:Vfinal-simp}) has the symmetry of the permutation group $S_3$ following from the underlying exchange symmetries ($1 \leftrightarrow 2 \leftrightarrow 3$) of the CY threefold. As we have discussed earlier, one can have isotropic moduli stabilisation if one appropriately choses the following exchange symmetry among the coefficients,
\bea
\label{eq:exchange-symmetry}
& & t^1 \leftrightarrow t^2 \leftrightarrow t^3, \quad {\cal C}_{w_1} \leftrightarrow {\cal C}_{w_2} \leftrightarrow {\cal C}_{w_3}, \quad {\cal C}_{w_4} \leftrightarrow {\cal C}_{w_5} \leftrightarrow {\cal C}_{w_6}.
\eea
The main idea is that the overall volume modulus ${\cal V}$ can be fixed using perturbative LVS while the other two moduli can effectively drive two-field-assisted fiber inflation by exploiting these underlying symmetries.

\subsection{Two-field assisted Fibre inflation}
Given that overall volume modulus (${\cal V}$) is stabilized at the leading order, we will work in the real basis $\Phi^a = \{\mathcal{V}, t^2, t^3\}$ of three K\"ahler moduli. The leading order contributions to the field space metric can be subsequently computed in the new basis using the tree-level K\"ahler potential as below
\bea
K_{T_\alpha\ov{T}_\beta} (\partial_\mu T_\alpha) (\partial^\mu \ov{T}_\beta) = \frac{1}{2} {\cal G}_{ab} (\partial_\mu \Phi^a) \,(\partial^\mu\Phi^b),
\eea
which gives
\bea
\label{eq:metric-phia}
& & \hskip-1.5cm {\cal G}_{ab} = \left(\begin{array}{ccc}
\frac{1}{{\cal V}^2} &  -\frac{1}{2 t^2 {\cal V}} & -\frac{1}{2 t^3 {\cal V}} \\
 -\frac{1}{2 t^2 {\cal V}} & \frac{1}{(t^2)^2} & \frac{1}{2 t^2 t^3} \\
 -\frac{1}{2 t^3 {\cal V}} &  \frac{1}{2 t^2 t^3} & \frac{1}{(t^3)^2} \\
\end{array}
\right), \qquad {\cal G}^{ab} = \left(\begin{array}{ccc}
\frac{3{\cal V}^2}{2} & \frac{{\cal V}t^2}{2} & \frac{{\cal V}t^3}{2} \\
 \frac{{\cal V}t^2}{2} & \frac{3 (t^2)^2}{2} & -\frac{t^2t^3}{2} \\
 \frac{{\cal V}t^3}{2} &  -\frac{t^2t^3}{2}  & \frac{3 (t^3)^2}{2}\\
\end{array}
\right).
\eea
Subsequently, the non-vanishing Christoffel connections $\Gamma^a_{bc}$ are given as below,
\bea
\label{eq:Christofell}
& & \Gamma_{11}^1 = - \frac{1}{\cal V}, \qquad \Gamma_{22}^2 = - \frac{1}{t^2}, \qquad \Gamma_{33}^3 = - \frac{1}{t^3}.
\eea
Now, we consider the full scalar potential in the real field basis $\Phi^a = \{\mathcal{V}, t^2, t^3\}$ which is:
\begin{align}
V(\mathcal{V}, t^2, t^3) &= \frac{\mathcal{C}_{\rm up}}{\mathcal{V}^p} + \frac{\mathcal{C}_1}{\mathcal{V}^3}\left(\hat{\xi} + 2\hat{\eta}\ln\mathcal{V} - 8\hat{\eta} + 2\hat{\sigma}\right) \\
&- \frac{\mathcal{C}_2}{\mathcal{V}^3}\left(2\mathcal{C}_{w_1}\frac{t^2t^3}{\mathcal{V}} + \frac{\mathcal{C}_{w_2}}{t^2} + \frac{\mathcal{C}_{w_3}}{t^3} + \frac{\mathcal{C}_{w_4}t^2t^3}{\mathcal{V} + 2(t^2)^2t^3} + \frac{\mathcal{C}_{w_5}}{2(t^2+t^3)} + \frac{\mathcal{C}_{w_6}t^2t^3}{\mathcal{V} + 2t^2(t^3)^2}\right) \nonumber\\
&+ \frac{\mathcal{C}_3}{\mathcal{V}^3}\left(\frac{1}{2t^2t^3} + \frac{t^2}{\mathcal{V}} + \frac{t^3}{\mathcal{V}}\right) + \cdots \nonumber
\end{align}
This potential exhibits exchange symmetry $t^2 \leftrightarrow t^3$ and depends on the uplifting mechanism characterized by the $p$ parameter such that $p=2$ corresponds to D-term uplifting, $p=8/3$ corresponds to T-brane uplifting, and $p=4/3$ corresponds to $\overline{D3}$-brane uplifting.

Using the  model-specific data, we can now explicitly write the second-order differential equations that describe the evolutionary dynamics.
 These are given as follows,
\bea
\label{eq:Explicit-EOMs}
& & \hskip-1cm {\cal V}^{\prime\prime} = \frac{{\cal V}^{\prime2}}{{\cal V}} - \left(3- \epsilon_H\right) \left({\cal V}^{\prime}+\frac{3{\cal V}^2}{2V} \partial_{\cal V} V +\frac{{\cal V} \, t^2}{2V}\partial_{t^2} V +\frac{{\cal V} \, t^3}{2V}\partial_{t^3} V \right),\\
& & \hskip-1cm (t^2)^{\prime\prime} = \frac{(t^2)^{\prime2}}{t^2} - \left(3- \epsilon_H\right) \left((t^2)^{\prime}+\frac{{\cal V} \, t^2}{2V}\partial_{{\cal V}} V +\frac{3(t^2)^2}{2V} \partial_{t^2} V -\frac{t^2 t^3}{2V}\partial_{t^3} V\right),\nonumber\\
& & \hskip-1cm (t^3)^{\prime\prime} = \frac{(t^3)^{\prime2}}{t^3} - \left(3- \epsilon_H\right) \left((t^3)^{\prime}+\frac{{\cal V} \, t^3}{2V}\partial_{{\cal V}} V -\frac{t^2 t^3}{2V}\partial_{t^2} V +\frac{3(t^3)^2}{2V} \partial_{t^3} V \right).\nonumber
\eea
Here as earlier, the prime $^\prime$ denotes derivatives with respect to the number of $e$-folds $N$, i.e. ${\cal V}^\prime = \frac{d{\cal V}}{dN}$ etc, and now the inflationary parameter $\epsilon_H$ takes the following explicit form,
\bea
& & \hskip-1cm \epsilon_H = \frac{1}{2} \left(\frac{{\cal V}^{\prime2}}{{\cal V}^2} + \frac{(t^2)^{\prime2}}{(t^2)^2}  + \frac{(t^3)^{\prime2}}{(t^3)^2} - \frac{{\cal V}^\prime \, (t^2)^{\prime}}{{\cal V} \, t^2} - \frac{{\cal V}^\prime \, (t^3)^{\prime}}{{\cal V} \, t^3} + \frac{(t^2)^\prime \, (t^3)^{\prime}}{t^2\, t^3} \right).
\eea

\subsubsection*{Numerical Analysis}

\noindent
In order to determine the evolutionary trajectories of various fields, we now solve the second-order differential equation~(\ref{eq:Explicit-EOMs}) under the following initial conditions:
\bea
& & \Phi^a(0)=\Phi^a_0 \qquad  {\rm and} \qquad \frac{d\Phi^a}{dN}|_{N=0}=0\,.\eea
After solving the evolution equations numerically, we have presented several benchmark models compatible with observational data with and/or without incorporating the recent ACT+DESI results \cite{Leontaris:2025hly}. For demonstration purposes, here we present one of those models \cite{Leontaris:2025hly}:
\bea
\label{eq:model-M3-3-ACT}
& & \hskip-1.5cm p = 8/3, \qquad \chi({\rm CY}) = -224, \qquad \eta_0 = 6, \qquad \sigma_0 = -4, \qquad g_s = 0.295, \\
& & \hskip-1.5cm  |W_0| = 5, \qquad {\cal C}_{w_1} = 0.001, \qquad  {\cal C}_{w_2} = -0.0008, \qquad  {\cal C}_{w_3} = -0.0008,    \nonumber\\
& & \hskip-1.5cm {\cal C}_{w_4} = -0.1, \qquad {\cal C}_{w_5} = 0.33, \qquad  {\cal C}_{w_6} = -0.1, \qquad \lambda = - 0.00017; \nonumber\\
& & \nonumber\\
& & \hskip-1.5cm {\cal C}_{\rm up} = 5.32455, 
\quad \langle {\cal V} \rangle = 1123.23, \quad \langle t^2 \rangle = 1.14996, \quad \langle t^3 \rangle = 1.14996, \nonumber\\
& & \hskip-1.5cm \langle \phi^1 \rangle = 6.01802, \quad \langle \phi^2 \rangle = -2.41341  , \quad \langle \phi^3 \rangle = -2.41341 , \nonumber\\
& & \nonumber\\
& & \hskip-1.5cm {\cal V}^\ast = 1258.22, \qquad (t^2)^\ast = 25.8, \quad  (t^3)^\ast = 25.8, \nonumber\\
& & \hskip-1.5cm \phi^{1\ast} = 6.11069, \quad \phi^{2\ast} = 1.35 , \quad \phi^{3\ast} = 1.35, \quad \Delta\phi = 5.32, \quad N = 55.5, \quad N^\ast = 5.5,\nonumber\\
& & \hskip-1.5cm  P_s^\ast =  2.095\cdot 10^{-9}, \quad n_s^\ast = 0.9763, \quad \alpha_s^\ast = -5.763\cdot10^{-4}, \quad r^\ast = 2.73\cdot10^{-3}.\nonumber
\eea
The effective inflaton shift $\Delta\phi$ can be considered as the distance in the flat three-dimensional field space spanned by canonical fields ${\phi^a}$, and it turns out that
\bea
\label{eq:shiftphi-three-field}
& & \Delta \phi = \sqrt{\left(\phi^{1\ast} - \langle\phi^1\rangle\right)^2 + \left(\phi^{2\ast} - \langle\phi^2\rangle\right)^2 + \left(\phi^{3\ast} - \langle\phi^3\rangle\right)^2}~,
\eea
which is simply the Pythagorean distance between two points. Here, $\phi^{a\ast}$ corresponds to the horizon exit while $\langle \phi^a\rangle$ denotes the moduli VEVs at the perturbative LVS minimum. In fact, for any choice of canonical field basis, the inflaton shift $\Delta\phi$ remains invariant. However, the individual fields may result in different shifts in a different choice of basis. Since our scalar potential has exchange symmetry $2 \leftrightarrow 3$ inherited from the CY threefold, we have used the following basis of canonical fields $\phi^a$
\bea
\label{eq:cononical-varphi3}
& & \phi^1 = \frac{1}{\sqrt{3}} \left(\varphi^1+ \varphi^2 + \varphi^3 \right) = \sqrt{\frac{2}{3}} \ln(\sqrt{2}\,{\cal V}) , \\
& & \phi^2 = \frac{1}{6} \left(2\sqrt{3}\, \varphi^1 + (3 -\sqrt{3})\varphi^2-(3+\sqrt{3} \phi^3) \right), \nonumber\\
& & \phi^3 = \frac{1}{6} \left(2\sqrt{3}\, \varphi^1 - (3 +\sqrt{3})\varphi^2 + (3 - \sqrt{3}) \phi^3 \right),\nonumber
\eea
where 
\bea
\label{eq:cononical-varphi0}
& & \varphi^a = \frac{1}{\sqrt{2}} \ln \tau_a, \qquad \forall \, a \in \{1, 2, 3\}.
\eea
For the benchmark model presented in (\ref{eq:model-M3-3-ACT}), we have
\begin{equation}
\Delta\phi^1 \simeq 0.0926, \qquad \Delta\phi^2 = \Delta\phi^3 \simeq 3.763 \qquad \Rightarrow \qquad \Delta\phi \simeq 5.32.
\end{equation}
This illustrates a key advantage of assisted inflation: while single-field models require $\Delta\phi \simeq 5.3 M_p$, the two-field approach reduces individual excursions to be around $3.7 M_p$, mitigating concerns about trans-Planckian displacements.

\section{Tetra-quadric CY Threefold}
\label{sec_tetra-quadric}

As we have demonstrated in the minimal assisted fibre inflation model with two K\"ahler moduli, an underlying symmetry in the Calabi-Yau geometry is crucial. Given that perturbative LVS fixes one of the K\"ahler moduli in the form of the overall volume of the Calabi-Yau threefold while rest of the moduli may assist during the inflationary process, one would need the Calabi-Yau threefolds with $h^{1,1}({\rm CY}) = 4$ for an effective three-field fibre inflation model. The motivation for this extension is to investigate the possibility of reducing the individual inflaton shifts while still reproducing the values of the various cosmological observables consistent with the experimental results.

Further, we also note that multiple K3-fibrations help us for the purpose of having the desired underlying symmetry in the CY threefold which can facilitate the ``assisted" nature of the inflationary dynamics. For example, the volume form of the previous model
\bea
{\cal V} = 2 \, t^1 t^2 t^3 = t^1 \tau_1 = t^2 \tau_2 = t^3 \, \tau_3,
\eea
follows from the fact that the only non-trivial triple intersection number turns out to be $k_{123} = 2$, which subsequently leads to an exchange symmetry $1 \leftrightarrow 2 \leftrightarrow 3$ in the CY volume form, divisors volumes, and also takes it through the scalar potential pieces as well. An extension of such a symmetry needs to be implemented for Calabi-Yau threefold with $h^{1,1}({\rm CY}) = 4$, say by generalizing the Intersection polynomial. While scanning the (Kreutzer-Skarke) Calabi-Yau database we find that Tetra-quadric hypersurface realized in ${\mathbb P}^2 \times {\mathbb P}^2 \times {\mathbb P}^2 \times {\mathbb P}^2$ turns out to possess an exchange symmetry $1 \leftrightarrow 2 \leftrightarrow 3 \leftrightarrow 4$ in the CY volume form. We take this observation as an initial motivation for using Tetra-quadric CY for our three-field fibre inflation purpose. The main idea is to consider type IIB superstring compactification on an orientifold of the tetra-quadric CY threefold where the overall volume is stabilized via perturbative LVS by the leading-order effects while the subleading corrections depending on the volume of the other three K3 divisors drive inflation.

\subsection{Relevant topological data}
For the illustration purpose of finding the multi K3-fibred CY with underlying exchange symmetry, we begin with a so-called tetra-quadric CY threefold. This corresponds to the polytope Id: 681 in the AGHJN database of \cite{Altman:2014bfa,Altman:2021pyc}, and can be simply defined by the following toric data:
\begin{center}
\begin{tabular}{|c|cccccccc|}
\hline
\cellcolor[gray]{0.9}Hyp &\cellcolor[gray]{0.9} $x_1$  &\cellcolor[gray]{0.9} $x_2$  &\cellcolor[gray]{0.9} $x_3$  &\cellcolor[gray]{0.9} $x_4$  &\cellcolor[gray]{0.9} $x_5$ & \cellcolor[gray]{0.9}$x_6$  &\cellcolor[gray]{0.9} $x_7$ &\cellcolor[gray]{0.9} $x_8$      \\
\hline
\cellcolor[gray]{0.9}2 & 0  & 0 & 0 & 1 & 1 & 0  & 0 & 0  \\
\cellcolor[gray]{0.9}2 & 0  & 0 & 1 & 0 & 0 & 1  & 0 & 0 \\
\cellcolor[gray]{0.9}2 & 0  & 1 & 0 & 0 & 0 & 0  & 1 & 0 \\
\cellcolor[gray]{0.9}2 & 1  & 0 & 0 & 0 & 0 & 0  & 0 & 1 \\
\hline
& K3  & K3 & K3 &  K3 & K3 & K3 & K3 & K3  \\
\hline
\end{tabular}
\end{center}
The Hodge numbers are $(h^{2,1}, h^{1,1}) = (68, 4)$, the Euler number is $\chi=-128$ while the Stanley-Reisner ideal reads as:
\be
{\rm SR} =  \{x_1 x_8, \, x_2 x_7, \, x_3 x_6, \, x_4 x_5 \} \,. \nn
\ee
This CY threefold also appears in the complete intersection Calabi-Yau (CICY) database \cite{Green:1986ck,Candelas:1987kf,Green:1987cr,Gray:2014fla,Anderson:2017aux} which has a total of 7880 CY threefolds. These CY threefolds are realized in product of projective spaces ${\mathbb P}^n$, e.g. the current CY is realized as a hypersurface in ${\mathbb P}^2 \times {\mathbb P}^2 \times {\mathbb P}^2 \times {\mathbb P}^2$. The analysis of the divisor topologies using {\it cohomCalg} \cite{Blumenhagen:2010pv, Blumenhagen:2011xn} shows that all of the coordinate divisors are K3 surfaces which have the following Hodge numbers:
\bea
& & \hskip-1.5cm {\rm K3} \equiv \begin{tabular}{ccccc}
    & & 1 & & \\
   & 0 & & 0 & \\
  1 & & 20 & & 1 \\
   & 0 & & 0 & \\
    & & 1 & & \\
  \end{tabular}. \nonumber
\eea
Considering the basis of smooth divisors $\{D_1, D_2, D_3, D_4\}$ we get the following intersection polynomial,
\bea
& & I_3 = 2\, \hat{D}_1\, \hat{D}_2\, \hat{D}_3 + 2\, \hat{D}_1\, \hat{D}_2\, \hat{D}_4 + 2\, \hat{D}_1\, \hat{D}_3\, \hat{D}_4 + 2\, \hat{D}_2\, \hat{D}_3\, \hat{D}_4,
\eea
while the second Chern-class of the CY is given by,
\bea
c_2({\rm CY}) = 4 \hat{D}_1 \hat{D}_2 + 4 \hat{D}_1 \hat{D}_3 + 4 \hat{D}_1 \hat{D}_4 + 4 \hat{D}_2 \hat{D}_3 + 4 \hat{D}_2 \hat{D}_4 + 4 \hat{D}_3 \hat{D}_4.
\eea
Subsequently, the second Chern numbers, defined as $\Pi_\alpha = \int_{\rm CY} c_2({\rm CY}) \wedge \hat{D}_\alpha$ corresponding to the divisors $D_\alpha$, are given as,
\bea
& & \Pi_\alpha = 24 \quad \forall \, \alpha \in \{1, 2,..,8\}.
\eea
Moreover, considering the K\"ahler form $J = \sum\limits_{\alpha =1}^4 t^\alpha \hat{D}_\alpha$, the overall volume ${\cal V}$ can be given as follows,
\bea
& & {\cal V} = 2\, t^1\, t^2\, t^3 + 2\, t^1\, t^2\, t^4 + 2\, t^1\, t^3\, t^4 +2\, t^2\, t^3\, t^4,
\eea
while the 4-cycle volume moduli, $\tau_\alpha = \partial_\alpha {\cal V}$, corresponding to the K3 divisors are given as below
\bea
& & \hskip-1cm \tau_1 = 2\, t^2 t^3 + 2\, t^2 t^4 + 2\, t^3 t^4,  \qquad  \tau_2 = 2\, t^1 t^3 + 2\, t^1 t^4 + 2\, t^3 t^4, \nonumber\\
& & \hskip-1cm \tau_3 = 2 \,t^1 t^2 + 2 \,t^1 t^4+ 2 \,t^2 t^4, \qquad \tau_4 = 2 \,t^1 t^2 + 2 \,t^1 t^3+ 2 \,t^2 t^3 \,. \nonumber
\label{Taus}
\eea
Now, unlike the previous example, it is hard to express the overall volume form in terms of $\tau_\alpha$'s only. Nevertheless it may be worth mentioning the following relations:
\bea
& & \hskip-1.5cm t^1 \tau_1 = {\cal V} - 2 \, t^2\, t^3\, t^4, \quad t^2 \tau_2 = {\cal V} - 2 \, t^1\, t^3\, t^4, \quad t^3 \tau_3 = {\cal V} - 2 \, t^1\, t^2\, t^4, \quad t^4 \tau_4 = {\cal V} - 2 \, t^1\, t^2\, t^3,
\eea
which satisfy the general relation: $t^\alpha \tau_\alpha = 3 {\cal V}$. Thus the volume form ${\cal V}$ inherits  an exchange symmetry $1 \leftrightarrow 2 \leftrightarrow 3 \leftrightarrow 4$ under which all the four K3 divisors which are part of the basis are exchanged. Further, the K\"ahler cone for this setup is described by the conditions below,
\bea
\label{KahCone}
& & \hskip-1.5cm \text{K\"ahler cone:} \quad  t^1 > 0\,, \quad t^2 > 0\,, \quad t^3 > 0\,, \quad t^4 > 0\,.
\eea
The intersection curves between two coordinate divisor are presented in Table \ref{Tab3} where similar to the toroidal case, the only non-trivial intersection among the K3-divisors are 2-torus.
\begin{table}[h]
  \centering
 \begin{tabular}{|c|c|c|c|c|c|c|c|c|}
\hline
\cellcolor[gray]{0.9}  &\cellcolor[gray]{0.9} $D_1$  &\cellcolor[gray]{0.9} $D_2$  &\cellcolor[gray]{0.9} $D_3$  & \cellcolor[gray]{0.9}$D_4$  & \cellcolor[gray]{0.9}$D_5$ &\cellcolor[gray]{0.9} $D_6$  & \cellcolor[gray]{0.9}$D_7$ & \cellcolor[gray]{0.9}$D_8$ \\
    \hline
		\hline
\cellcolor[gray]{0.9}$D_1$ & $\emptyset$  &  ${\mathbb T}^2$      &  ${\mathbb T}^2$        &  ${\mathbb T}^2$   &  ${\mathbb T}^2$  &  ${\mathbb T}^2$   &  ${\mathbb T}^2$ & $\emptyset$ \\
\cellcolor[gray]{0.9}$D_2$ & ${\mathbb T}^2$  &  $\emptyset$      &  ${\mathbb T}^2$        &  ${\mathbb T}^2$   &  ${\mathbb T}^2$  &  ${\mathbb T}^2$   &  $\emptyset$ & ${\mathbb T}^2$
\\
\cellcolor[gray]{0.9}$D_3$  & ${\mathbb T}^2$  &  ${\mathbb T}^2$      &  $\emptyset$        &  ${\mathbb T}^2$   &  ${\mathbb T}^2$  &  $\emptyset$   &  ${\mathbb T}^2$ & ${\mathbb T}^2$ \\
\cellcolor[gray]{0.9}$D_4$  & ${\mathbb T}^2$  &  ${\mathbb T}^2$      &  ${\mathbb T}^2$        &  $\emptyset$   &  $\emptyset$  &  ${\mathbb T}^2$   &  ${\mathbb T}^2$ & ${\mathbb T}^2$ \\
\cellcolor[gray]{0.9}$D_5$ & ${\mathbb T}^2$  &  ${\mathbb T}^2$      &  ${\mathbb T}^2$        &  $\emptyset$   &  $\emptyset$  &  ${\mathbb T}^2$   &  ${\mathbb T}^2$ & ${\mathbb T}^2$ \\
\cellcolor[gray]{0.9}$D_6$ & ${\mathbb T}^2$  &  ${\mathbb T}^2$      &  $\emptyset$        &  ${\mathbb T}^2$   &  ${\mathbb T}^2$  &  $\emptyset$   &  ${\mathbb T}^2$ & ${\mathbb T}^2$ \\
\cellcolor[gray]{0.9}$D_7$ & ${\mathbb T}^2$  &  $\emptyset$      &  ${\mathbb T}^2$        &  ${\mathbb T}^2$   &  ${\mathbb T}^2$  &  ${\mathbb T}^2$   &  $\emptyset$ & ${\mathbb T}^2$ \\
\cellcolor[gray]{0.9}$D_8$ & $\emptyset$  &  ${\mathbb T}^2$      &  ${\mathbb T}^2$        &  ${\mathbb T}^2$   &  ${\mathbb T}^2$  &  ${\mathbb T}^2$   &  ${\mathbb T}^2$ & $\emptyset$ \\
\hline
  \end{tabular}
  \caption{Intersection curves of the two coordinate divisors.}
\label{Tab3}
\end{table}
Subsequently, using the K\"ahler form $J = t^1 \hat{D}_1 + t^2 \hat{D}_2 +t^3 \hat{D}_3 + t^4 \hat{D}_4$ the corresponding sizes of the curves in Table \ref{Tab3} can be expressed in terms of two-cycle volumes $t^\alpha$ as presented in Table \ref{Tab4}. Also, this table is symmetrical and lower left entries can be read-off from the upper right block.
\begin{table}[h]
  \centering
 \begin{tabular}{|c|c|c|c|c|c|c|c|c|}
\hline
 \cellcolor[gray]{0.9} &\cellcolor[gray]{0.9} $D_1$  &\cellcolor[gray]{0.9} $D_2$  & \cellcolor[gray]{0.9}$D_3$  &\cellcolor[gray]{0.9} $D_4$  &\cellcolor[gray]{0.9} $D_5$ & \cellcolor[gray]{0.9}$D_6$  &\cellcolor[gray]{0.9} $D_7$ &\cellcolor[gray]{0.9} $D_8$ \\
    \hline
		\hline
\cellcolor[gray]{0.9}$D_1$ & 0 & 2 ${t^3}$+2 ${t^4}$ & 2 ${t^2}$+2 ${t^4}$ & 2 ${t^2}$+2 ${t^3}$ & 2 ${t^2}$+2 ${t^3}$ & 2 ${t^2}$+2
   ${t^4}$ & 2 ${t^3}$+2 ${t^4}$ & 0 \\
\cellcolor[gray]{0.9}$D_2$ &  & 0 & 2 ${t^1}$+2 ${t^4}$ & 2 ${t^1}$+2 ${t^3}$ & 2 ${t^1}$+2 ${t^3}$ & 2 ${t^1}$+2
   ${t^4}$ & 0 & 2 ${t^3}$+2 ${t^4}$ \\
\cellcolor[gray]{0.9}$D_3$ &  &  & 0 & 2 ${t^1}$+2 ${t^2}$ & 2 ${t^1}$+2 ${t^2}$ & 0 & 2
   ${t^1}$+2 ${t^4}$ & 2 ${t^2}$+2 ${t^4}$ \\
\cellcolor[gray]{0.9}$D_4$ &  &  &  & 0 & 0 & 2 ${t^1}$+2 ${t^2}$ & 2
   ${t^1}$+2 ${t^3}$ & 2 ${t^2}$+2 ${t^3}$ \\
\cellcolor[gray]{0.9}$D_5$  &  &  &  &  & 0 & 2 ${t^1}$+2 ${t^2}$ & 2
   ${t^1}$+2 ${t^3}$ & 2 ${t^2}$+2 ${t^3}$ \\
\cellcolor[gray]{0.9}$D_6$  &  &  &  &  &  & 0 & 2
   ${t^1}$+2 ${t^4}$ & 2 ${t^2}$+2 ${t^4}$ \\
\cellcolor[gray]{0.9}$D_7$  &  &  &  &  &  &  & 0 & 2 ${t^3}$+2 ${t^4}$ \\
\cellcolor[gray]{0.9}$D_8$  &  &  &  & &  &  &  & 0 \\
\hline
  \end{tabular}
  \caption{Size of the curves at the intersection locus of the two  divisors presented in Table \ref{Tab3}. }
\label{Tab4}
\end{table}
\noindent


\subsection{Orientifold involution, fluxes and brane setting}
For a given holomorphic involution, one needs to introduce D3/D7-branes and fluxes in order to cancel all the charges. For example, one can nullify the D7-tadpoles via introducing stacks of $N_a$ D7-branes wrapped around suitable divisors (say $D_a$) and their orientifold images ($D_a^\prime$) such that the following relation holds \cite{Blumenhagen:2008zz}:
\bea
\label{eq:D7tadpole}
& & \sum_a\, N_a \left([D_a] + [D_a^\prime] \right) = 8\, [{\rm O7}]\,.
\eea
Moreover, the presence of D7-branes and O7-planes also contributes to the D3-tadpoles, which, in addition, receive contributions from  $H_3$ and $F_3$ fluxes, D3-branes and O3-planes. The D3-tadpole cancellation condition is given as \cite{Blumenhagen:2008zz}:
\be
N_{\rm D3} + \frac{N_{\rm flux}}{2} + N_{\rm gauge} = \frac{N_{\rm O3}}{4} + \frac{\chi({\rm O7})}{12} + \sum_a\, \frac{N_a \left(\chi(D_a) + \chi(D_a^\prime) \right) }{48}\,,
\label{eq:D3tadpole}
\ee
where $N_{\rm flux} = (2\pi)^{-4} \, (\alpha^\prime)^{-2}\int_X H_3 \wedge F_3$ is the contribution from background fluxes and $N_{\rm gauge} = -\sum_a (8 \pi)^{-2} \int_{D_a}\, {\rm tr}\, {\cal F}_a^2$ is due to  worldvolume fluxes from D7. However, for the simple case where D7-tadpoles are canceled by placing 4 D7-branes (plus their images) on top of an O7-plane, equation (\ref{eq:D3tadpole}) reduces to the following form:
\be
N_{\rm D3} + \frac{N_{\rm flux}}{2} + N_{\rm gauge} =\frac{N_{\rm O3}}{4} + \frac{\chi({\rm O7})}{4}\,.
\label{eq:D3tadpole1}
\ee
For this CY threefold, we note that there are eight equivalent reflection involutions corresponding to flipping eight coordinates, i.e. $x_i \to - x_i$ for each $i \in \{1, 2, .., 8\}$. The details of the respective Fixed points are summarized in Table \ref{tab_FixedPointSet}.

\begin{table}[H]
\centering
\hskip0.11cm \begin{tabular}{|c|c||c|c|c|}
\hline
Sr. No. & Involution & O7-planes & O3-planes  & $Q_{D3}$ \\
\hline
1 & $\sigma_1$ & $\{D_1, D_8\}$ & $\emptyset$ & 24 \\
2 & $\sigma_2$ & $\{D_2, D_7\}$ & $\emptyset$ & 24 \\
3 & $\sigma_3$ & $\{D_3, D_6\}$ & $\emptyset$ & 24 \\
4 & $\sigma_4$ & $\{D_4, D_5\}$ & $\emptyset$ & 24 \\
5 & $\sigma_5$ & $\{D_4, D_5\}$ & $\emptyset$ & 24 \\
6 & $\sigma_6$ & $\{D_3, D_6\}$ & $\emptyset$ & 24 \\
7 & $\sigma_7$ & $\{D_2, D_7\}$ & $\emptyset$ & 24 \\
8 & $\sigma_8$ & $\{D_1, D_8\}$ & $\emptyset$ & 24 \\
\hline
\end{tabular}
\caption{Fixed point set for a given involution $\sigma_i: x_i \to - x_i$. Here, the D3 tadpole charge $Q_{\rm D3}$ corresponds to the brane setting where D7-branes are placed on top of the O7-planes.}
\label{tab_FixedPointSet}
\end{table}

\subsection{Scalar potential with various (sub-)leading effects}
Now we discuss the various types of (sub-)leading string-loop corrections which can possibly be induced in this setting. Subsequently, the scalar potential of interest can be expressed as,
\bea
\label{eq:potV1+V2}
& & V \equiv V({\cal V}, t^2, t^3, t^4) = V_1({\cal V}) + V_2({\cal V}, t^2, t^3, t^4),
\eea
where 
\bea
& & \hskip-1cm V_1({\cal V}) = \frac{{\cal C}_{\rm up}}{{\cal V}^p} + V_{\rm pLVS}, \qquad V_2({\cal V}, t^2, t^3, t^4) = V_{g_s}^{\rm KK} + V_{g_s}^{\rm W} + V_{{\rm F}^4}.
\eea
The first piece $V_1({\cal V})$ appears at leading order with typically a dependence on the overall volume ${\cal V}$ only, and it keeps the rest of the moduli still flat due to the underlying no-scale structure. This piece fixes the overall volume modulus via the perturbative LVS. Subsequently, with a suitable choice of the Winding parameters ${\cal C}_{w_\alpha}$, the second piece $V_2$ can effectively be considered as a three-field potential suitable for assisted fibre inflation due to a symmetry $2 \leftrightarrow 3 \leftrightarrow 4$ following form the underlying CY threefold.

\subsubsection*{BBHL and log-loop corrections at leading order}
The leading order no-scale breaking contributions are induced through the so-called BBHL's ${\alpha^\prime}^3$ corrections \cite{Becker:2002nn} along with the log-loop effects \cite{Antoniadis:2018hqy,Antoniadis:2018ngr,Antoniadis:2019doc}. As we mentioned earlier, leading order contributions to this effect is encoded in the following pieces,
\bea
& & V_{\rm pLVS} = \frac{{\cal C}_1}{{\cal V}^3} \left(\hat\xi + 2\,\hat\eta \, \ln{\cal V} - 8\,\hat\eta + 2\,\hat\sigma \right).
\eea

\subsubsection*{Winding-type string loop corrections}
For the simple involutions $\sigma_i$ which reflect a single toric coordinate $x_i$, \i.e., $\sigma_i: x_i \to - x_i$ we observe that there are no non-intersecting D7-brane stacks with non-shrinkable intersections loci. This subsequently means that there should be no Winding-type contributions a la prescription of \cite{Berg:2004ek, Berg:2005ja, Berg:2005yu, Berg:2007wt, Cicoli:2007xp}, however, as argued in \cite{vonGersdorff:2005bf,Gao:2022uop,Gao:2026mvc}, generically one can still expect some Winding-type corrections. Having that in mind, we consider the generic possibilities for this model by looking at the volumes of the curves lying at the intersection of all the eight coordinate divisors. From Table~\ref{Tab3}, we also observe that any two divisors either do not intersect or intersect on a two-torus. Considering the K\"ahler form $J \equiv t^\alpha \hat{D}_\alpha = t^1 \hat{D}_1 + t^2 \hat{D}_2 + t^3 \hat{D}_3 + t^4 \hat{D}_4$, subsequently one has
\bea
& & t_{\cap}\left(D_\alpha \cap D_\beta\right) \equiv \int_{\rm CY} J \wedge \hat{D}_\alpha \wedge \hat{D}_\beta = k_{\alpha\beta\gamma}\, t^\gamma,
\eea
which results in Table \ref{Tab4}, and we also recall that the non-trivial triple intersection numbers always take value 2, i.e. $\{k_{123} = 2, \, k_{124} = 2, \, k_{134} = 2, \, k_{234} = 2\}$. This leads to the following generic Winding-type contributions which may be possible to the scalar potential,
\bea
\label{eq:Vgs-Winding-globalmodel-2a}
& & \hskip-0.75cm V_{g_s}^{\rm W} \equiv - \frac{2\kappa\,|W_0|^2}{{\cal V}^3} \, \sum_{\alpha=1}^6 \frac{C_\alpha^W}{t_\cap^\alpha} \\ 
& & = - \frac{\kappa\,|W_0|^2}{{\cal V}^3} \, \biggl[\frac{C_1^W}{(t^1+t^2)} + \frac{C_2^W}{(t^1+t^3)} + \frac{C_3^W}{(t^1+t^4)} + \frac{C_4^W}{(t^2+t^3)} + \frac{C_5^W}{(t^2+t^4)} + \frac{C_6^W}{(t^3+t^4)}\biggr]. \nonumber
\eea

\subsubsection*{KK-type string loop corrections}
The KK-type corrections should be absent for the chosen class of holomorphic involutions which leads to no O3-planes and O7-planes have same line bundle charges resulting in on-top situation. However, the generic argument irrespective of the specific brane-setting may allow for the following contributions of KK-type string loop corrections,
\bea
\label{eq:Vkktype}
& & \hskip-1cm V_{g_s}^{\rm KK} = \frac{g_s^2\, |W_0|^2\, \kappa}{16\, {\cal V}^4} \biggl[({\cal C}_1^{\rm KK})^2 \biggl(\frac{4{\cal V}^2}{\tau_1^2} + \frac{2}{\tau_1^2} \biggl\{\left(t^3-t^4\right)^2 \left(t^3+t^4\right) \left(t^2\right)^3\\
& & -t^3 t^4 \left(\left(t^3\right)^2-4 t^4 t^3+\left(t^4\right)^2\right) \left(t^2\right)^2 -\left(t^3\right)^2 \left(t^4\right)^2 \left(t^3+t^4\right) t^2+\left(t^3\right)^3 \left(t^4\right)^3 \biggr\}\biggr) \nonumber\\
& & + ({\cal C}_2^{\rm KK})^2\,\left(\tau_1 + 4\, (t^2)^2\right) + ({\cal C}_3^{\rm KK})^2\,\left(\tau_1 + 4\, (t^3)^2\right) + ({\cal C}_4^{\rm KK})^2\,\left(\tau_1 + 4\, (t^4)^2\right) \biggr]\,, \nonumber
\eea
where $\tau_1 = 2(t^2 t^3 + t^2 t^4 + t^3 t^4)$. However, we do not aim to consider such corrections in the current analysis.

\subsubsection*{Higher derivative F$^4$ corrections}
Finally, although this CY has several properties similar to the toroidal case, the divisor being K3 implies their corresponding $\Pi({\rm K3})= 24$, and therefore, unlike the ${\mathbb T}^4$ case which has a vanishing $\Pi$, there are following higher derivative F$^4$ corrections to the scalar potential,
\bea
\label{eq:F^4-term-globalmodel}
& & V_{{\rm F}^4} = - \frac{\lambda\,\kappa^2\,W_0^4}{g_s^{3/2} {\cal V}^4}\, 24 \, \left(t^1 + t^2 + t^3 + t^4\right).
\eea


\section{Moduli Stabilization}
\label{sec_moduli-stab}
In this section, we discuss moduli stabilisation in perturbative LVS, and subsequently explore the three-field dynamics for the assisted fibre inflation while considering the overall volume modulus to be fixed at leading order.

\subsection{Symmetries of the scalar potential and (almost) isotropic vacua}
The final scalar potential is a collection of the following contributions,
\bea
\label{eq:scalarpotential-gen}
& & \hskip-1cm V({\cal V}, t^2, t^3, t^4) \simeq \frac{{\cal C}_{\rm up}}{{\cal V}^p} + \frac{{\cal C}_1}{{\cal V}^3} \left(\hat\xi + 2\,\hat\eta \, \ln{\cal V} - 8\,\hat\eta + 2\,\hat\sigma \right) \\
& & - \frac{{\cal C}_2}{{\cal V}^3} \, \biggl[\frac{C_1^W}{(t^1+t^2)} + \frac{C_2^W}{(t^1+t^3)} + \frac{C_3^W}{(t^1+t^4)} + \frac{C_4^W}{(t^2+t^3)} + \frac{C_5^W}{(t^2+t^4)} + \frac{C_6^W}{(t^3+t^4)}\biggr]\nonumber\\
& & + \frac{{\cal C}_3}{{\cal V}^4}\,\left(t^1 + t^2 + t^3 + t^4\right) + \dots, \nonumber
\eea
where one needs to eliminate $t^1 \equiv t^1({\cal V}, t^2, t^3, t^4)$ as
\bea
& & t^1 = \frac{{\cal V} - 2 t^2 t^3 t^4}{2\left(t^2 t^3 + t^2 t^4 + t^3 t^4\right)}.
\eea
As we discussed earlier, the pieces in the first line which only depend on the overall volume ${\cal V}$ fix it in the perturbative AdS LVS minimum uplifted to de-Sitter minimum, while the pieces in the second and third line stabilize the rest of the three moduli. Due to the presence of the residual $S_3$ permutation symmetry arising from  the underlying $S_4$ symmetry of the CY threefold, one can look for `isotropic' solutions for these three moduli by demanding the following constraint in the choice of model dependent parameters appearing in the Winding-type string loop correction,
\bea
\label{eq:condCwsIso}
& & C_1^W = C_2^W = C_3^W \equiv \tilde{\cal C}_w, \qquad C_4^W = C_5^W = C_6^W \equiv {\cal C}_w.
\eea
For a (partial) isotropic moduli stabilization following from the constrained choice of winding parameters in (\ref{eq:condCwsIso}), the scalar potential (\ref{eq:scalarpotential-gen}) can be split into two pieces: one that determines the minimum and another that gives rise to the steepening. The parameters must then be chosen so as to produce a sufficiently long plateau. In this form, the potential can be re-expressed as
\bea
\label{}
& & \hskip-1cm V({\cal V}, t^2, t^3, t^4) = V_{\rm up} + V_{\rm pLVS} + V_{W}^{(i)} + V_{W}^{(ii)} + V_{{\rm F}^4}^{(i)} + V_{{\rm F}^4}^{(ii)},
\eea
where
\bea
& & V_{\rm up} = \frac{{\cal C}_{\rm up}}{{\cal V}^p}, \qquad V_{\rm pLVS} = \frac{{\cal C}_1}{{\cal V}^3} \left(\hat\xi + 2\,\hat\eta \, \ln{\cal V} - 8\,\hat\eta + 2\,\hat\sigma \right),\\
& & V_{W}^{(i)} = - \frac{{\cal C}_2}{{\cal V}^3} \, \left(\frac{{\cal C}_w}{(t^2+t^3)} + \frac{{\cal C}_w}{(t^2+t^4)} + \frac{{\cal C}_w}{(t^3+t^4)}\right), \quad V_{{\rm F}^4}^{(i)} = \frac{{\cal C}_3}{2\, {\cal V}^3\,(t^2 t^3 + t^2  t^4 + t^3 t^4)},\nonumber\\
& & V_{W}^{(ii)} = - \frac{2 \, {\cal C}_2\,(t^2 t^3 + t^2  t^4 + t^3 t^4)}{{\cal V}^4} \biggl[\frac{\tilde{\cal C}_w}{1 + \frac{2(t^2+t^3)(t^4)^2}{{\cal V}}} + \frac{\tilde{\cal C}_w}{1 + \frac{2(t^2+t^4)(t^3)^2}{{\cal V}}} + \frac{\tilde{\cal C}_w}{1 + \frac{2(t^3+t^4)(t^2)^2}{{\cal V}}}\biggr],\nonumber\\
& & V_{{\rm F}^4}^{(ii)} = \frac{{\cal C}_3\, (t^2+t^3)(t^2+t^4)(t^3+t^4)}{(t^2 t^3 + t^2  t^4 + t^3 t^4)\, {\cal V}^4}, \nonumber
\eea
where the pieces in the first two lines determine the minimum while the pieces in the last  two lines induce the steepening part arising from the winding-loop and F$^4$ corrections. 

Subsequently, it turns out that the ${\cal V}$ is fixed by the perturbative LVS while the Winding-type string loop corrections and the higher derivative F$^4$ corrections stabilize the remaining three moduli with the following respective VEVs corresponding to an AdS solution to be uplifted to a de-Sitter solution,
\bea
\label{eq:ta-VeV}
& & \langle{\cal V}\rangle \simeq e^{\frac{13}{3}-\frac{\hat\xi}{2\, \hat\eta} -\frac{\hat\sigma}{ \hat\eta}}, \qquad \langle t^a \rangle \simeq \frac{2{\cal C}_3}{9{\cal C}_2\, {\cal C}_w} = - \frac{8\, |W_0|^2\lambda}{3\, {\cal C}_w \sqrt{g_s}}, \qquad a \in \{2, 3, 4\}\,;
\eea
where the above mentioned leading order estimates may receive some ${\cal O}({\cal V}^{-1})$ corrections. Given that $\lambda$ typically receives a negative contribution, one needs to consider ${\cal C}_w > 0$ for the model building purpose while looking at the relevant region of the parameter space. Further, the (diagonal pieces in the) Hessian take the following VEV,
\bea
\label{eq:Hess-Vtata}
& & \langle V_{t^a t^a} \rangle = \frac{243 \, {\cal C}_2^4\, {\cal C}_w^4}{16\, {\cal C}_3^3\, \langle{\cal V}\rangle^3} + \frac{2{\cal C}_2\, {\cal C}_w}{3 \langle{\cal V}\rangle^4} + {\cal O}\left(\frac{1}{\langle{\cal V}\rangle^5}\right) \simeq -\frac{9\, {\cal C}_w^4 g_s^{5/2}}{2048\, |W_0|^4 \lambda^3\, \langle{\cal V}\rangle^3} + \frac{g_s\, {\cal C}_w \, |W_0|^2}{3\, \langle{\cal V}\rangle^4},
\eea
ensuring that the leading order piece is positive for $\lambda < 0$. Further, the VEV of the second piece of the potential $V_2$ which depends on the $\langle t^a\rangle $ moduli for $a \in \{2, 3, 4\}$,
\bea
\label{eq:VeV-pot-iso}
& & \langle V_2({\langle{\cal V}\rangle, t^2, t^3, t^4})\rangle = \frac{9{\cal C}_w^2 g_s^{3/2}}{64 \, \lambda \, \langle{\cal V}\rangle^3} + \frac{64\, (2{\cal C}_w - 3\, \tilde{\cal C}_w) |W_0|^6\, \lambda^2}{3 {\cal C}_w^2 \langle{\cal V}\rangle^4} + {\cal O}\left(\frac{1}{\langle {\cal V}\rangle^5}\right)~.
\eea
This shows that the minimum is actually an AdS minimum for  $\lambda < 0$, which requires a further slight shift to increase it to a de Sitter minimum, beyond what is needed to compensate the AdS minimum of the leading-order of the perturbative LVS.
It should be noted that our discussion of iterative or step-wise moduli stabilisation is intended only to facilitate the analytic understanding. In practice, our numerical approach fixes all four moduli simultaneously by considering the full potential $V = V_1 + V_2$ as given in (\ref{eq:scalarpotential-gen}).

On this occasion, it is important to mention that the mere volume scaling can sometime lead to a misleading outcome as there are several model dependent parameters involved that can collectively become competitive to the ${\cal V}$ factor even for their ${\cal O}(1)$ or small values. This is a point that requires careful attention when building a numerical model. However, for the sake of an analytic understanding of underlying dynamics with a complicated four- or three-field potential, it is worth comparing the leading order expressions.

\subsection{Two benchmark models}

Using the scalar potential (\ref{eq:scalarpotential-gen}) we perform four-field moduli stabilization. To illustrate the analytic discussion above, we now present a couple of numerical models.

\subsubsection*{Model A}
\bea 
\label{eq:modelA}
& & \chi({\rm CY}) = -128, \qquad p = 8/3, \qquad \eta_0 = 8, \qquad \sigma_0 = -6,\\
& & g_s = 0.24, \qquad W_0 = 2.3, \qquad {\cal C}_w = 0.13, \qquad \tilde{\cal C}_w = - 0.001, \qquad \lambda = - 0.001 \nonumber\\
& & {\cal C}_{\rm up} = 0.481727, \qquad \langle{\cal V}\rangle = 959.714, \qquad \langle{t^a}\rangle = 0.22148; \quad \forall a \in \{2, 3, 4\},\nonumber\\
& & \nonumber\\
& & \langle V \rangle \simeq 8.4786\cdot10^{-24}, \quad \langle V_{\rm up} \rangle \simeq 5.37554\cdot10^{-9}, \quad \langle V_{\rm pLVS} \rangle \simeq  -5.05949\cdot10^{-9}, \nonumber\\
& & \langle V_W^{(i)} \rangle \simeq -6.32282\cdot10^{-10}, \qquad \langle V_W^{(ii)} \rangle \simeq 6.60682\cdot10^{-16},\nonumber\\
& & \langle V_{{\rm F}^4}^{(i)} \rangle \simeq 3.1617\cdot10^{-10}, \qquad \quad \langle V_{{\rm F}^4}^{(ii)} \rangle \simeq 5.7267\cdot10^{-14}. \nonumber 
\eea

\subsubsection*{Model B}
\begin{figure}[H]
\centering
\includegraphics[width=16.5cm]{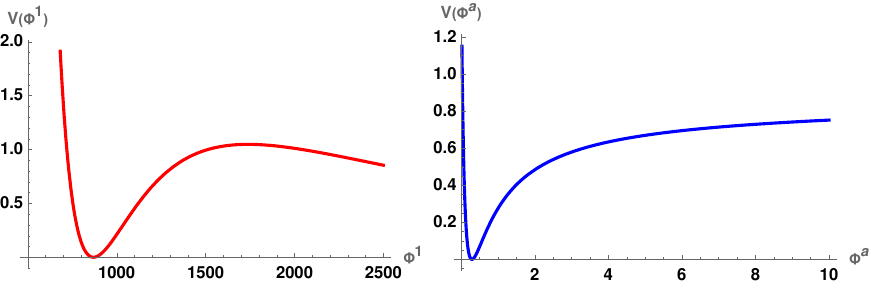}
\caption{Scalar potential ($V\cdot 10^{10}$) plotted for a single modulus at a time while assuming the other moduli to be fixed at their respective minimum. Here $\Phi^1$ is the overall volume ${\cal V}$ while $\Phi^a = \{t^2, t^3, t^4\}$ for $a =\{1, 2, 3\}$ due to the underlying exchange symmetry, namely $2 \leftrightarrow 3 \leftrightarrow 4$.}
\label{fig_V1}
\end{figure}

\begin{figure}[H]
\centering
\includegraphics[width=16.5cm]{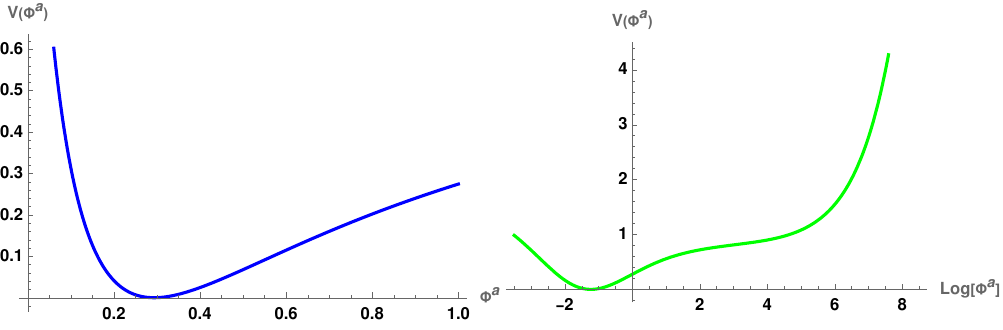}
\caption{Scalar potential ($V\cdot 10^{10}$) plotted for the individual inflaton moduli $\Phi^a = \{t^2, t^3, t^4\}$ while assuming the overall volume ${\cal V}$ to be fixed at its respective minimum. The right side figure manifestly shows the typical single-field flat track of the fibre inflation model.}
\label{fig_V2}
\end{figure}

\begin{figure}[H]
\centering
\includegraphics[width=16.9cm]{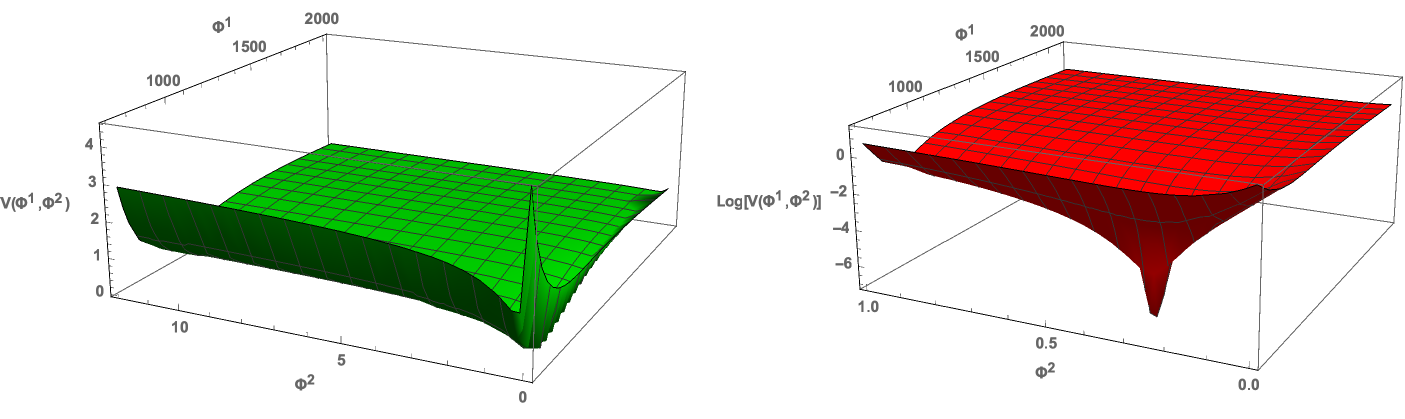}
\caption{Two dimensional plot of the scalar potential ($V\cdot 10^{10}$) while assuming the $\{t^3, t^4\}$ moduli to be fixed at their respective minimum. The second figure plotted for ($\ln(V\cdot 10^{10})$) clearly demonstrates the minimum as compared to the first one.}
\label{fig_V3}
\end{figure}

\bea
\label{eq:modelB}
& & \chi({\rm CY}) = -128, \qquad p = 8/3, \qquad \eta_0 = 8, \qquad \sigma_0 = -6,\\
& & g_s = 0.29, \qquad W_0 = 4.0, \qquad {\cal C}_w = 0.027, \qquad \tilde{\cal C}_w = - 0.0043, \qquad \lambda = - 0.0001 \nonumber\\
& & {\cal C}_{\rm up} = 1.99996, \qquad \langle{\cal V}\rangle = 869.547, \quad \langle{t^a}\rangle = 0.293344; \quad \forall a \in \{2, 3, 4\},\nonumber\\
& & \nonumber\\
& & \langle V \rangle \simeq 1.70606\cdot10^{-24}, \quad \langle V_{\rm up} \rangle \simeq 2.90339\cdot10^{-8}, \quad  \langle V_{\rm pLVS} \rangle \simeq  -2.87905\cdot10^{-8}, \nonumber\\
& & \langle V_W^{(i)} \rangle \simeq -4.87177\cdot10^{-10}, \qquad \langle V_W^{(ii)} \rangle \simeq 2.70246\cdot10^{-14},\nonumber\\
& & \langle V_{{\rm F}^4}^{(i)} \rangle \simeq 2.43672\cdot10^{-10}, \qquad \quad \langle V_{{\rm F}^4}^{(ii)} \rangle \simeq 1.13178\cdot10^{-13}. \nonumber 
\eea
These two benchmark models illustrate the following features:
\begin{itemize}
\item{The moduli stabilization is performed with all the four-fields considered together, and subsequently the uplifting piece takes care of not only the leading negative contributions from the perturbative LVS but also those arising from the sub-leading effects, namely the Winding loop corrections and the higher derivative corrections.}
\item{The following hierarchy is clearly manifested at the perturbative LVS minimum
\bea
& & \left|\langle V_{\rm pLVS}\rangle\right| \gg \left|\langle V_{g_s}^{\rm W}\rangle\right| \gtrsim \left|\langle V_{{\rm F}^4}\rangle\right|\,,
\eea
which means that there is a clear hierarchy between the piece which stabilize the overall volume ${\cal V}$ and those which drive inflation.}
\item{We also observe a clear hierarchy (for the VEVs) among the pieces which stabilize the volume moduli and those which create steepening as seen from
\bea
\left|\langle V_W^{(i)} \rangle\right| \simeq \left|\langle V_{{\rm F}^4}^{(i)}\right| \gg \left|\langle V_W^{(ii)} \rangle\right| \simeq \left|\langle V_{{\rm F}^4}^{(ii)}\right|
\eea
}
\item{As we will see later, {\bf Model A} produces the standard cosmological observables consistent with PLANCK data \cite{Planck:2018jri,Planck:2018vyg} while {\bf Model B} is the ACTivated version \cite{ACT:2025tim,ACT:2025fju,DESI:2024mwx, Frolovsky:2025iao} with a slighty larger value of the spectral $n_s$ and a non-negative value of the running of spectral index $\alpha_{s}$.}  

\item{For illustration purpose we present plots of the scalar potential in terms of varying a given set of fields, one at a time, in Fig.~\ref{fig_V1}-Fig.~\ref{fig_V3}.}

\end{itemize}

\section{Assisted Fibre Inflation}
\label{sec_assisted-FI}
In this section, we present the numerical analysis leading to a fibre inflation model assisted by three alike moduli, namely $\Phi^a=\{t^2, t^3, t^4\}$, supported by a residual exchange symetry ($2 \leftrightarrow 3 \leftrightarrow 4$) arising from the underlying symmetry of the Calabi-Yau threefold.

\subsection{Numerical analysis of inflationary dynamics}
Using the (inverse-)metric and Christoffel connections given in (\ref{eq:Gab})-(\ref{eq:affine}), the explicit form of the field equations for $\Phi^a=\{{\cal V}, t^2, t^3, t^4\}$ can be obtained from Eq.~(\ref{eq:EOM2}). These are given as
\bea
\label{eq:Explicit-EOMs}
& & \hskip-1cm {\cal V}^{\prime\prime} - \frac{{\cal V}^{\prime2}}{{\cal V}} + \left(3- \epsilon_H\right) \left({\cal V}^{\prime}+\frac{3{\cal V}^2}{2V} \partial_{\cal V} V +\frac{{\cal V} \, t^2}{2V}\partial_{t^2} V +\frac{{\cal V} \, t^3}{2V}\partial_{t^3} V +\frac{{\cal V} \, t^4}{2V}\partial_{t^4} V \right) = 0,\\
& & \hskip-1cm (t^2)^{\prime\prime} + \Gamma^2_{ab} {\Phi^a}^\prime\, {\Phi^b}^\prime + \left(3- \epsilon_H\right) \left((t^2)^{\prime}+\frac{{\cal G}^{2a}\, \left(\partial_a V\right)}{V}\right) = 0,\nonumber\\
& & \hskip-1cm (t^3)^{\prime\prime} + \Gamma^3_{ab} {\Phi^a}^\prime\, {\Phi^b}^\prime + \left(3- \epsilon_H\right) \left((t^3)^{\prime}+\frac{{\cal G}^{3a}\, \left(\partial_a V\right)}{V}\right) = 0,\nonumber\\
& & \hskip-1cm (t^4)^{\prime\prime} + \Gamma^4_{ab} {\Phi^a}^\prime\, {\Phi^b}^\prime + \left(3- \epsilon_H\right) \left((t^4)^{\prime}+\frac{{\cal G}^{4a}\, \left(\partial_a V\right)}{V}\right) = 0.\nonumber
\eea
Here, the prime $^\prime$ denotes derivatives w.r.t.~the number of $e$-folds $N$, i.e. ${\cal V}^\prime = \frac{d{\cal V}}{dN}$ etc. and the inflationary parameter $\epsilon_H$ takes the following explicit form,
\bea
& & \hskip-1cm \epsilon_H \equiv \frac{1}{2} \, {\cal G}_{ab} \frac{d\Phi^a}{dN} \frac{d\Phi^b}{dN} \simeq \frac{1}{2} \biggl[\frac{{\cal V}^{\prime2}}{{\cal V}^2} + \frac{4(t^3+t^4)^2}{(\tau_1)^2} \, (t^2)^{\prime2} + \frac{4(t^2+t^4)^2}{(\tau_1)^2} \, (t^3)^{\prime2} + \frac{4(t^2+t^3)^2}{(\tau_1)^2} \, (t^4)^{\prime2}\\
& & - \frac{2(t^3+t^4)}{\tau_1\,{\cal V}} \, {\cal V}^\prime \, (t^2)^{\prime} - \frac{2(t^2+t^4)}{\tau_1\,{\cal V}} \, {\cal V}^\prime \, (t^3)^{\prime} - \frac{2(t^2+t^3)}{\tau_1\,{\cal V}} \, {\cal V}^\prime \, (t^4)^{\prime} \nonumber\\
& & + \left(\frac{2}{\tau_1}+ \frac{8(t^4)^2}{\tau_1^2}\right) \, (t^2)^\prime \, (t^3)^{\prime} + \left(\frac{2}{\tau_1}+ \frac{8(t^3)^2}{\tau_1^2}\right) \, (t^2)^\prime \, (t^4)^{\prime} + \left(\frac{2}{\tau_1}+ \frac{8(t^2)^2}{\tau_1^2}\right) \, (t^3)^\prime \, (t^4)^{\prime}\biggr]. \nonumber
\eea
Note that the underlying $S_4$ symmetry (namely $1 \leftrightarrow 2 \leftrightarrow 3 \leftrightarrow 4$) inherited from the tetra-quadric CY threefold is reduced to a $S_3$ symmetry (with $2 \leftrightarrow 3 \leftrightarrow 4$) for the field-space basis $\Phi^a = \{{\cal V}, t^2, t^3, t^4\}$, and this is reflected in the set of four-field equations. 

Now one has to numerically solve the couple second order evolution equations under the following constraints
\bea
& & \Phi^a(0)=\Phi^a_0 \qquad  {\rm and} \qquad \frac{d\Phi^a}{dN}|_{N=0}=0\,.\eea
Given the extremely complicated nature of the metric, inverse metric and the affine connection, it is hard to perform a four-field analysis. However, we can assume that the overall volume ${\cal V}$ which is stabilized at the leading order perturbative LVS sits at its minimum during the inflationary process and does not contribute significantly in generating the efolds. In fact, earlier multi-field analysis have shown this to happen as long as a hierarchy between the pieces which stabilize the overall volume and those which drive inflation can be maintained. We perform a three-field analysis for the fields $\{t^2, t^3, t^4\}$ and the cosmological observables for the two benchmark models presented earlier are given in Table \ref{tab_cosmo-observables}.

\begin{table}[H]
\centering
\begin{tabular}{|c||c|c|c|c|c||c|c|c|c|} 
 \hline
& & & & & & & & & \\
{\bf Model} & ${t^a}^\ast$ & $N_{\rm tot}$ & $N^\ast$ & $\epsilon_H^\ast$ & $\eta_H^\ast$ & $P_s^\ast$ & $n_S^\ast$ & $\alpha_{n_S}^\ast$ & $r^\ast$\\ 
& & & & & & ($\times 10^9$) & & & \\
\hline
& & & & & & & & & \\
{\bf A} & 9.15 & 54 & 4 & 0.000595 & 0.032537 & 2.14 & 0.966266 & -0.000810 & 0.009523 \\
& & & & & & & & & \\
{\bf B} & 12.7 & 54 & 3 & 0.000465 & 0.025739 & 2.19 & 0.973328 & 0.000156 & 0.007437\\
& & & & & & & & & \\
 \hline
\end{tabular}
\caption{Cosmological Observables evaluated at the Horizon exit $N^\ast$.}
\label{tab_cosmo-observables}
\end{table}

\subsection{Tracking various evolutions through efolds}
In this subsection, we present the graphical version of the evolutions and track the scalar potential, inflaton field, and various cosmological observables in terms of the efolds. We present the figures only for the ACTivated model, i.e. {\bf Model B}, and those corresponding figures of {\bf Model A} happen to be similar.

\begin{figure}[H]
\centering
\includegraphics[width=15.1cm]{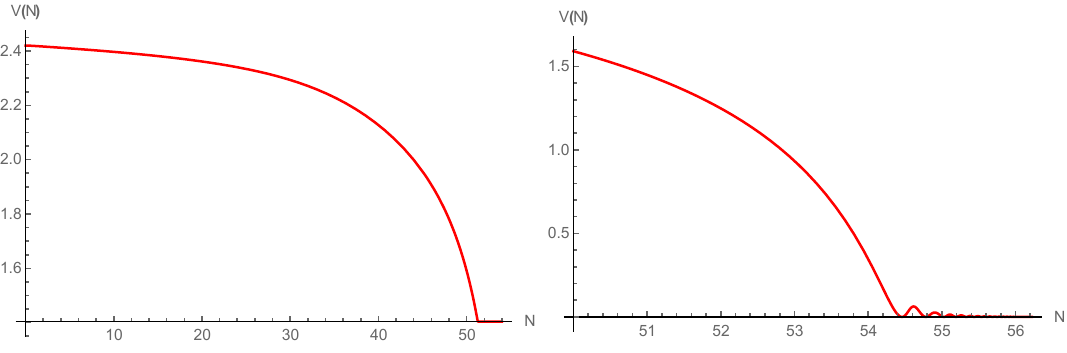}
\caption{Evolution of the scalar potential $V(N)\cdot10^{10}$ plotted for the number of efoldings}
\label{fig_pot-N-three-field}
\end{figure}

\begin{figure}[H]
\centering
\includegraphics[width=15.1cm]{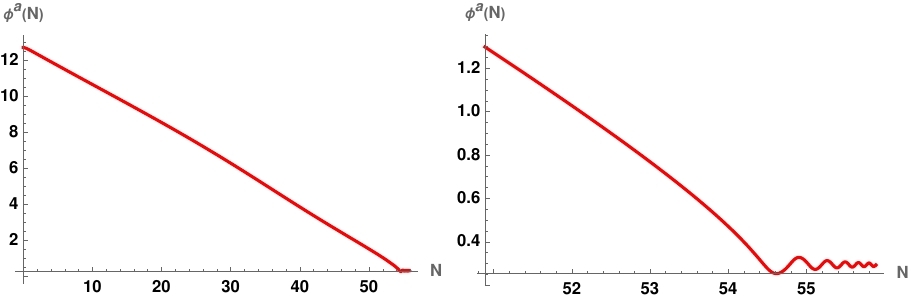}
\caption{Evolution of $\Phi^a(N)$ for $a = \{2, 3, 4\}$}
\label{fig_phi1-three-field}
\end{figure}

\begin{figure}[H]
\centering
\includegraphics[width=15.1cm]{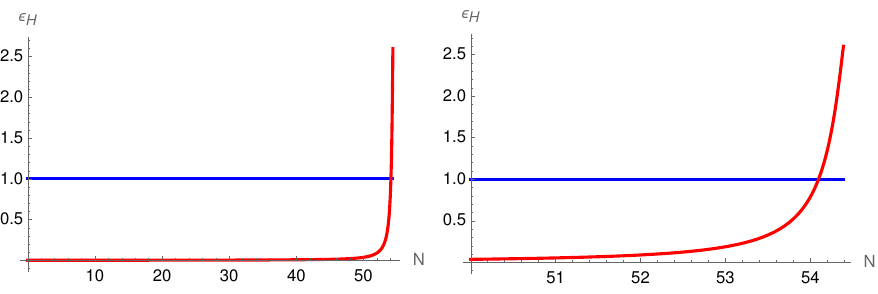}
\caption{Evolution of slow-roll parameter $\epsilon_H(N)$}
\label{fig_epsilon-three-field}
\end{figure}

\begin{figure}[H]
\centering
\includegraphics[width=15.1cm]{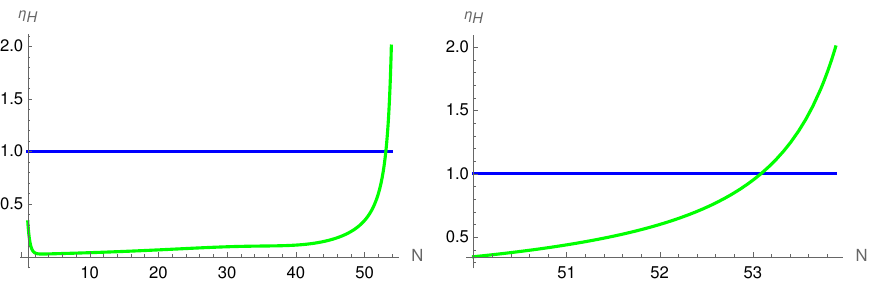}
\caption{Evolution of slow-roll parameter $\eta_H(N)$}
\label{fig_eta-three-field}
\end{figure}

\begin{figure}[H]
\centering
\includegraphics[width=16.1cm]{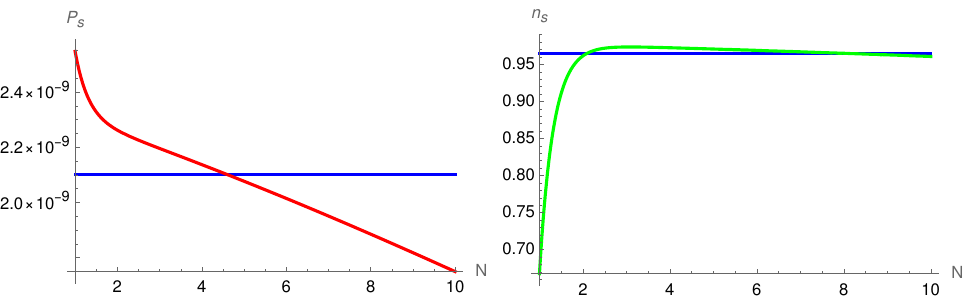}
\caption{Evolution of  $P_s(N)\cdot10^9$ and $n_s(N)$ with dashed lines for $P_s = 2.1 \cdot 10^{-9}$ and $n_s = 0.975$.}
\label{fig_Ps-ns-three-field}
\end{figure}

\begin{figure}[h!]
\centering
\includegraphics[width=16.1cm]{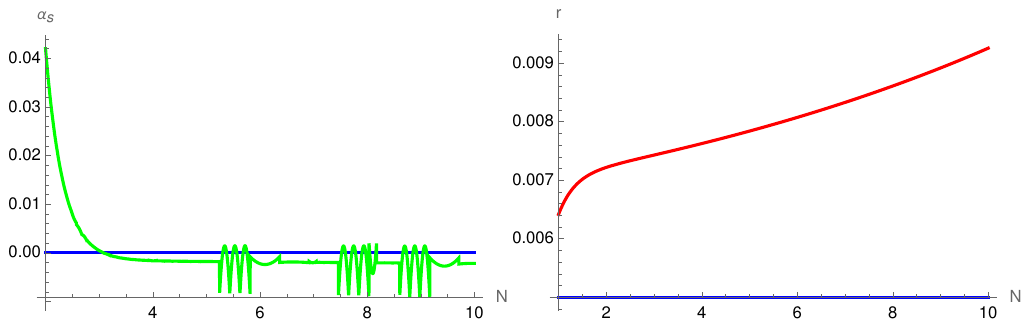}
\caption{Evolution of $\alpha_s(N)$ and  $r(N)$ 
using definitions in (\ref{eq:cosmo-observables1}) and $\phi\equiv\phi(N)$.}
\label{fig_r-three-field}
\end{figure}

\subsection{Inflaton field range and assisted nature of inflation}
Given the moduli space metric ${\cal G}_{ab}$ for the fields $\Phi^a$, the distance is given as
\bea
& & ds^2 = {\cal G}_{ab} \, d\Phi^a \, d\Phi^b,
\eea
and subsequently the length of a curve $\Phi^a = \Phi^a(\gamma)$ parameterized by the parameter $\gamma$ is given as,
\bea
& & \Delta\Phi = \int_{\gamma_1}^{\gamma_2} \, ds = \int_{\gamma_1}^{\gamma_2} \,\sqrt{{\cal G}_{ab} \frac{d\Phi^a}{d\gamma} \, \frac{d\Phi^b}{d\gamma}}\, d\gamma
\eea
For our purpose, given that we have already solved the inflationary trajectories in terms of the efolds $N$, one can take it as a good parameter to characterize the effective inflationary curve in the three-dimensional moduli space. Subsequently, the effective field excusrion during the entire inflationary process can be estimated as 
\bea
& & \Delta\Phi = \int_{N^\ast}^{N_{\rm end}} \,\sqrt{{\cal G}_{ab} \frac{d\Phi^a}{dN} \, \frac{d\Phi^b}{dN}}\, dN = \int_{N^\ast}^{N_{\rm end}} \,\sqrt{2\,\epsilon_H(N)}\, dN.
\eea
Now using this approach, we can not only compute the full effective inflaton shift but can also estimate the individual distances traveled by each of the three fields by considering the motion of one modulus while keeping the other two at their respective minima. Subsequently, we compare these ``distances" to have an estimate about the effect of ``assisted" nature of the inflation in the multi-field dynamics. The subsequent results are  presented in Table \ref{tab_inflaton-shifts}. 

\begin{table}[H]
\centering
\begin{tabular}{|c||c|c|c|} 
 \hline
& & & \\
{\bf Model}  & $\Delta\Phi$ & $\Delta\Phi^a$ for $a \in \{2, 3, 4\}$ & $\Delta\Phi/\sqrt3$\\ 
& (effective) & (individual) & (Pythagorean estimate) \\
\hline
& & & \\
{\bf A}  & 6.44923 & 2.48231 & 3.72346 \\ 
& & & \\
{\bf B}  & 5.70785 & 2.19695 & 3.29543 \\ 
& & & \\
 \hline
\end{tabular}
\caption{Effective and individual inflaton shifts needed during the full inflationary process}
\label{tab_inflaton-shifts}
\end{table}

\noindent
The two estimates presented in Table \ref{tab_inflaton-shifts} clearly demonstrate the assisted nature of the inflationary dynamics. In fact, it is more effective that the Pythagorean estimate one has made in \cite{Leontaris:2025hly} where it was suggested that if the total effective inflaton shift is $\Delta\Phi$, say for the single-field case, then the presence of $n$ inflatons facilitates successful inflation via merely making a shift of each of the individual inflatons by $\Delta\Phi/\sqrt{n}$. Having better assistance than the canonical case is because of the fact that the non-trivial off-diagonal terms in the metric contribute to the effective inflaton shift. This is further supported by the residual exchange symmetry, namely $2 \leftrightarrow 3 \leftrightarrow 4$ following from the underlying CY threefold.

\subsection{Comments on the stability of the assisted dynamics}

\begin{figure}[H]
\centering
\includegraphics[width=16.5cm]{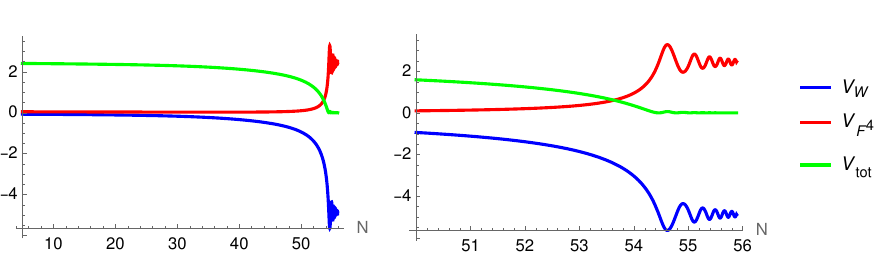}
\caption{Evolution of scalar potential ($V\cdot 10^{10}$) and its pieces showing that the effective inflaton potential receives negative contributions from the Winding-type corrections which is compensated by the F$^4$ corrections such that the total potential remains flat enough to drive inflation !}
\label{fig_V-scales1-three-field}
\end{figure}

For the illustration purposes, now we discuss the stability of the benchmark model {\bf Model B} by considering the various contributions to the scalar potential. Similar estimates should hold for the {\bf Model A} as well. It turns out that the perturbative AdS LVS minimum appears at the leading order which is appropriately uplifted to a de-Sitter minimum, and the individual scalar potential contributions are,
\bea
\label{eq:various-V-vevs1}
& & \hskip-1cm \langle V_{\rm up} \rangle = 2.90339 \cdot 10^{-8}, \qquad \langle V_{\rm pLVS} \rangle = -2.87905\cdot 10^{-8},\\
& & \hskip-1cm \langle V_{g_s}^W \rangle = -4.8715 \cdot 10^{-10}, \qquad \langle V_{{\rm F}^4} \rangle = 2.43785\cdot 10^{-10}. \nonumber
\eea
In turns out that during the inflationary process, i.e. significantly away from the minimum, the subleading contributions from Winding- and F$^4$-corrections are further suppressed, and therefore the overall volume ${\cal V}$ cannot receive any significant shift. The evolution of Winding-type and F$^4$ type corrections during the entire inflationary process is presented in figure \ref{fig_V-scales1-three-field}.

\noindent
In addition, the following eigenvalues of the mass-squared matrix indeed ensure a mass hierarchy among the overall volume modulus and the remaining three inflaton moduli 
\bea
\label{eq:mass-squared}
& & m_a^2 \simeq \{2.46\cdot 10^{-9}, \, \, 7.99\cdot 10^{-11}, \, \, 8.49 \cdot 10^{-14}, \, \, 8.49 \cdot 10^{-14}\},
\eea
where the heaviest eigenstate mostly corresponds to the overall volume modulus ${\cal V}$ while the other ones are some combinations of the remaining three moduli $\{t^2, t^3, t^4\}$. A clean mass hierarchy between the overall volume mode, e.g. see (\ref{eq:mass-squared}), and the three inflaton directions ensures that there can be only a very small shift in ${\cal V}$ modulus during inflation and the four-field evolution is effectively a three-field assisted inflationary dynamics as we have discussed.

For the convention in which string length $\ell_s$ and the $\alpha^\prime$ parameter are connected as $\ell_s = 2\pi \sqrt{\alpha^\prime}$ \cite{Conlon:2006gv}, the string mass $M_s$ and gravitino mass $m_{3/2}$ for {\bf Model B} are given as
\bea
\label{eq:KKscales}
& & M_s \equiv \frac{M_p}{\sqrt{\alpha^\prime}} = \frac{g_s^{1/4} \sqrt\pi}{\sqrt{\cal V}} M_p \simeq 4.4\cdot10^{-2}\, M_p, \\
& & m_{3/2} \equiv e^{\frac{1}{2} {K}} |W_0| = \frac{\sqrt{g_s}\, |W_0|}{\sqrt{2} \, {\cal V}} \simeq 1.75 \cdot 10^{-3} \, M_p, \nonumber
\eea
Furthermore, the various KK scales are defined as 
\bea
\label{eq:KKscales}
& & M_{\rm KK}^a = \frac{M_p}{R_a} = \frac{M_s}{\tilde{R}_a}\,,
\eea
where $R_a = \tilde{R}_a \sqrt{\alpha^\prime}$ where $\tilde{R}_a$ is the size of the relevant length corresponding to a particular KK mode such that $\tilde{R}_a = (t^a)^{1/2}$ for two-cycle volumes, $\tilde{R}_a = (\tau_a)^{1/4}$ for four-cycle volumes and $\tilde{R}_L = {\cal V}^{1/6}$  corresponds to the bulk modulus ($t^b \simeq{\cal V}^{1/3}$ or $\tau_b \simeq {\cal V}^{2/3}$) which we denote as $M_{\rm KK}^{\rm bulk}$. Usually these are the lightest KK modes. 

\begin{figure}[H]
\centering
\includegraphics[width=16.5cm]{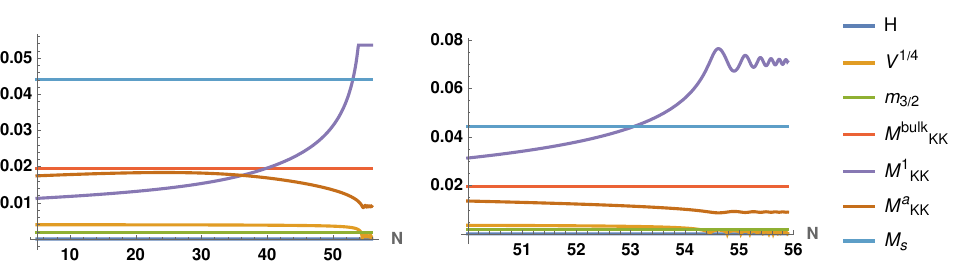}
\caption{Evolution of various mass scales during inflationary dynamics}
\label{fig_scales-three-field}
\end{figure}

Let us note that the following mass hierarchy should be respected during the inflationary evolution,
\bea
\label{eq:mass-hierarchy}
& & m_a < H < V^{1/4} < m_{3/2} < M_{\rm KK} < M_s < M_p,
\eea
where $m_a$ is the inflaton masses while $H$ is the Hubble scale. For {\bf Model B}, we have the following estimates at the horizon exit,
\bea
& & \hskip-1cm H^\ast \simeq 8.97 \cdot 10^{-6}, \quad M_{\rm KK}^{\rm bulk} \simeq 1.95\cdot 10^{-3}, \quad M_{\rm KK}^1 \simeq 1.11\cdot 10^{-2}, \quad M_{\rm KK}^{\rm a} \simeq 1.73\cdot 10^{-2},
\eea
The evolution of various mass scales are plotted in figure \ref{fig_scales-three-field} where apart from the bulk KK mode, we consider $M_{\rm KK}^1$ corresponding to $\tau_1$ modulus and $M_{\rm KK}^{a}$ corresponding to $\tau_2, \tau_3$ and $\tau_4$ moduli.

\noindent
The evolution of various scales as presented in figure \ref{fig_scales-three-field} shows that mass-hierarchy (\ref{eq:mass-hierarchy}) is respected throughout the inflationary regime, i.e. till $\epsilon_H \leq 1$ corresponding to $N\simeq54.08$. However, we also observe that one of the KK scales $M_{\rm KK}^{1}$ become comparable to the string mass towards the minimum, after the end of inflation, something which have been observed in \cite{Leontaris:2025hly} as well. However, as long as the volume moduli do not enter in the regime where they take too smaller values, the supergravity approximations should remain (marginally) valid, and the string mass scale being comparable to (one of the) KK mass may not be an immediate concern as argued in \cite{Dienes:2002ze}.


\section{Summary and conclusions}
\label{sec_conclusions}
In this article, we have demonstrated that multiple moduli fields can collectively assist in driving fibre inflation, effectively sharing the burden of generating the required number of e-folds. We began by reviewing the implementation of Fibre Inflation within the perturbative Large Volume Scenario (pLVS) in a multi K3-fibred Calabi-Yau threefold with $h^{1,1}=3$ and then extended our analysis to a tetra-quadric CY threefold featuring an underlying $S_4$ permutation symmetry. Considering a two-field analysis, it has been found in \cite{Leontaris:2025hly} that multi K3-fibration with an appropriate underlying symmetry in the CY threefold can help in reducing the individual canonical inflaton shift by a factor of $\sqrt2$. Subsequently it has been further argued that if the effective single-field inflaton shift needed for producing cosmological observables is $\Delta\Phi$, one can exploit using $n$ canonical inflaton fields $\Phi^a$ with some underlying symmetry such that the individual inflaton shifts are $\Delta\Phi^a = \Delta\Phi/\sqrt{n}$ as suggested by Pythagorean distance in the Euclidean space. Remarkably, in this extended setup of tetra-quadric CY threefold with $S_4$ exchange symmetry, we found that individual inflaton excursions can be further reduced while still reproducing cosmological observables in full agreement with current experimental data. We have substantiated this result with several illustrative benchmark models.

We performed a detailed analysis of the four-field moduli stabilization with various perturbative effects induced via the higher derivative ($\alpha^\prime$) and string-loop ($g_s$) corrections. The overall volume modulus is minimized by a combination of the so-called BBHL $\alpha'$  corrections and log-loop terms in the perturbative LVS, while the remaining three K\"ahler moduli collectively drive inflation via sub-leading string loop and higher-derivative corrections. The resulting construction successfully reproduces the observed cosmological observables, including the power spectrum amplitude $P_s$, spectral index $n_s$, tensor-to-scalar ratio $r$, and running of the spectral index $\alpha_{n_s}$, while maintaining all individual field displacements near $2.2$ M$_p$ although the total effective inflaton shift needed to be around $5.7$ M$_p$. Our work therefore establishes assisted inflation as a viable and theoretically consistent mechanism for realizing trans-Planckian inflation in string theory, simultaneously satisfying both observational constraints and quantum gravity consistency conditions. However, we also mention that there are some issues which need to be given further attention, e.g. we observe that the KK masses corresponding to the inflaton moduli get very close to the string mass towards the minimum. These observations suggest that overall EFT description may not be as clean and robust as one would like it to be.
	
This multi-field assisted inflation proposal suggests a new way to addresses a critical challenge in trans-Planckian string-inspired cosmology. In conventional fibre inflation models, the effective inflaton displacement is typically trans-Planckian, which poses serious problems: it pushes moduli toward the boundaries of the K\"ahler cone, at the risk of invalidating the effective field theory approximation, and may be excluded by swampland conjectures. Our proposal circumvents these difficulties by showing that the desired cosmological requirements can be satisfied in a suitably designed multi-field model without forcing individual inflaton fields too close to their respective K\"ahler cone boundaries. This is crucial because very large excursions would excite heavy moduli, breaking the mass hierarchy and undermining the decoupling assumption essential to the effective field theory framework which typically have only a few fields active in the low energy dynamics. On these lines, as a future plan, we aim to present a fibre inflation model with several moduli having a sub-Planckian shift in the individual inflaton.


\section*{Acknowledgments}
We are very thankful to Ignatios Antoniadis and Roberto Valandro for useful discussions. SB is grateful to the  {\it Department of Science and Technology} (DST), India, for providing the Junior Research Fellowship in the form of an {\it Institute Fellowship for PhD work} at Bose Institute. PS would like to gratefully acknowledge the {\it Mathematical Research Impact Centric Support {\rm (MATRICS)} grant} (Ref.~ANRF/ARGM/2025/002717/TS) received from the  Anusandhan National Research Foundation {(ANRF)}, India. In addition, PS would like to thank the {\it Department of Science and Technology} (DST), India for the kind support.


\appendix


\section{Moduli space metric and affine connections}
\label{sec_appendix}
The generic field space metric ${\cal G}_{ab}$ for a set of real moduli can be obtained from the K\"ahler moduli space metric $K_{T_\alpha\ov{T}_\beta}$ using the following relation
\bea
\label{eq:fieldspace-metric}
& & K_{T_\alpha\ov{T}_\beta} (\partial_\mu T_\alpha) (\partial^\mu \ov{T}_\beta) = \frac{1}{2} {\cal G}_{ab} (\partial_\mu \Phi^a) \,(\partial^\mu\Phi^b),
\eea
where $\{\Phi^a\}$ forms the new basis of the real fields. Subsequently, the leading order effects to the (non-flat) field space metric ${\cal G}_{ab}$ and the corresponding Christoffel connections $\Gamma^a_{bc}$ can be derived using the tree level K\"ahler potential, i.e. taking $K = - 2 \ln {\cal V}$ for ${\cal V} = 2 t^1 t^2 t^3 + 2 t^1 t^2 t^4 + 2 t^1 t^3 t^4 +2 t^2 t^3 t^4$. 

Considering the basis $\Phi^a = \{\tau_1, \tau_2, \tau_3, \tau_4\}$, where $\tau_\alpha$'s are four-cycle volumes corresponding to the four K3 divisors in the toric basis, one finds that the inverse metric ${\cal G}^{ab}$ can be given as below,
\bea
& & {\cal G}^{ab} = \left(\begin{array}{cccc}
\tau_1^2 & 4 (t^3)^2 (t^4)^2 & 4 (t^2)^2 (t^4)^2 & 4 (t^2)^2 (t^3)^2\\
& & \\
4 (t^3)^2 (t^4)^2 & \tau_2^2 & 4 (t^1)^2 (t^4)^2 & 4 (t^1)^2 (t^3)^2 \\
& & \\
4 (t^2)^2 (t^4)^2 & 4 (t^1)^2 (t^4)^2 & \tau_3^2 & 4 (t^1)^2 (t^2)^2 \\
& & \\
4 (t^2)^2 (t^3)^2 & 4 (t^1)^2 (t^3)^2 & 4 (t^1)^2 (t^2)^2 & \tau_4^2 \\
\end{array}
\right),
\eea
while the metric components ${\cal G}_{ab}$ are rather too complicated to collect here. This is because of the fact that the $t^\alpha \to \tau_\alpha$ conversion is highly non-trivial for the volume form ${\cal V}$ of the current CY threefold. This complication is further manifested in the complicated form of $k^{\alpha\beta} = \left(k_{\alpha\beta\gamma}.t^\gamma\right)^{-1}$

However, for our purpose let us recall that the more suitable basis to work with is $\Phi^a = \{{\cal V}, t^2, t^3, t^4\}$. This is due to the fact that the overall volume modulus ${\cal V}$ is stabilized by the leading order effects using the perturbative LVS, while the remaining three moduli are lifted by subleading effects which eventually also helps in realizing (an assisted version of fibre) inflation. Subsequently, the various components of the moduli space metric ${\cal G}_{ab}$ in the new basis is given as
\bea
\label{eq:Gab}
& & {\cal G}_{11} = \frac{1}{{\cal V}^2}, \qquad {\cal G}_{12} = -\frac{t^3 + t^4}{\tau_1\, {\cal V}}, \qquad {\cal G}_{13} = -\frac{t^2 + t^4}{\tau_1\, {\cal V}}, \qquad {\cal G}_{14} = -\frac{t^2 + t^3}{\tau_1\, {\cal V}},\\
& & {\cal G}_{22} = \frac{4(t^3 + t^4)^2}{\tau_1^2} +\frac{8(t^3 + t^4) (t^3)^2 (t^4)^2}{\tau_1^2\, {\cal V}}, \qquad {\cal G}_{23} = \frac{1}{\tau_1} + \frac{4 (t^4)^2}{\tau_1^2} - \frac{8 t^2 t^3 (t^4)^3}{\tau_1^2\, {\cal V}},\nonumber\\
& & {\cal G}_{24} = \frac{1}{\tau_1} + \frac{4 (t^3)^2}{\tau_1^2} - \frac{8 t^2 t^4 (t^3)^3}{\tau_1^2\, {\cal V}}, \qquad {\cal G}_{33} = \frac{4(t^2 + t^4)^2}{\tau_1^2} +\frac{8(t^2 + t^4) (t^2)^2 (t^4)^2}{\tau_1^2\, {\cal V}},\nonumber\\
& & {\cal G}_{34} = \frac{1}{\tau_1} + \frac{4 (t^2)^2}{\tau_1^2} - \frac{8 t^3 t^4 (t^2)^3}{\tau_1^2\, {\cal V}}, \qquad {\cal G}_{44} = \frac{4(t^2 + t^3)^2}{\tau_1^2} +\frac{8(t^2 + t^3) (t^2)^2 (t^3)^2}{\tau_1^2\, {\cal V}},\nonumber
\eea
where we have used $\tau_1 = 2 (t^2 t^3 + t^2 t^4 + t^3 t^4)$ just to save some space to fit the equations concisely. We note that the underlying exchange symmetry $2 \leftrightarrow 3 \leftrightarrow 4$ is clearly manifested in the moduli space metric. Let us also emphasize here that these expressions do not consider any large volume approximation so far, except the tree level assumption in the K\"ahler potential $K = - 2 \ln {\cal V}$. In the absence of the fourth modulus $t^4$ one has $\tau_1 \to 2 t^2 t^3$ and ${\cal V} \to 2 t^1 t^2 t^3$. Subsequently, one finds that the metric components get reduced to their respective expressions presented in the toroidal-like CY model in \cite{Leontaris:2025hly}.

For this metric ${\cal G}_{ab}$,  the inverse metric (${\cal G}^{ab}$) and the affine connections $(\Gamma^a_{bc})$ take rather some complicated forms and subsequently it is better to use the large volume expansion for presenting the various components by their respective leading order contributions. This is given as below,
\bea
\label{eq:InverseGab}
& & {\cal G}^{11} = \frac{3 {\cal V}^2}{2}, \qquad {\cal G}^{12} = \frac{{\cal V}\, t^2}{2}, \qquad {\cal G}^{13} = \frac{{\cal V} \,t^3}{2}, \qquad {\cal G}^{14} = \frac{{\cal V}\, t^4}{2},\\
& & {\cal G}^{22} \simeq \frac{3 (t^2)^2 + \tau_1}{2}, \qquad {\cal G}^{23} \simeq \frac{3 t^2\, t^3 - \tau_1}{2}, \qquad {\cal G}^{24} \simeq \frac{3 t^2\, t^4 - \tau_1}{2},\nonumber\\
& & {\cal G}^{33} \simeq \frac{3 (t^3)^2 + \tau_1}{2}, \qquad {\cal G}^{34} \simeq \frac{3 t^3\, t^4 - \tau_1}{2}, \qquad {\cal G}^{44} \simeq \frac{3 (t^4)^2 + \tau_1}{2},\nonumber
\eea
where the components in the first line are exact while the others can receive ${\cal O}({\cal V}^{-1})$ corrections. Finally, the affine connections $\Gamma^c_{ab}$ are collected in the following matrices,
\bea
\label{eq:affine}
& & \Gamma^1_{ab} = \left(
\begin{array}{cccc}
 -\frac{1}{{\cal V}} & 0 & 0 & 0 \\
 0 & 0 & 0 & 0 \\
 0 & 0 & 0 & 0 \\
 0 & 0 & 0 & 0 \\
\end{array}
\right), \qquad \Gamma^2_{ab} \simeq \left(
\begin{array}{cccc}
 0 & 0 & 0 & 0 \\
 0 & -\frac{2 \left(t^3+t^4\right)}{{\tau_1}} & -\frac{t^4}{{\tau_1}} & -\frac{t^3}{{\tau_1}} \\
 0 & -\frac{t^4}{{\tau_1}} & 0 & \frac{t^2}{{\tau_1}} \\
 0 & -\frac{t^3}{{\tau_1}} & \frac{t^2}{{\tau_1}} & 0 \\
\end{array}
\right),\\
& & \Gamma^3_{ab} \simeq \left(
\begin{array}{cccc}
 0 & 0 & 0 & 0 \\
 0 & 0 & -\frac{t^4}{{\tau_1}} & \frac{t^3}{{\tau_1}} \\
 0 & -\frac{t^4}{{\tau_1}} & -\frac{2 \left(t^2+t^4\right)}{{\tau_1}} & -\frac{t^2}{{\tau_1}} \\
 0 & \frac{t^3}{{\tau_1}} & -\frac{t^2}{{\tau_1}} & 0 \\
\end{array}
\right), \qquad \Gamma^4_{ab} \simeq \left(
\begin{array}{cccc}
 0 & 0 & 0 & 0 \\
 0 & 0 & \frac{t^4}{{\tau_1}} & -\frac{t^3}{{\tau_1}} \\
 0 & \frac{t^4}{{\tau_1}} & 0 & -\frac{t^2}{{\tau_1}} \\
 0 & -\frac{t^3}{{\tau_1}} & -\frac{t^2}{{\tau_1}} & -\frac{2 \left(t^2+t^3\right)}{{\tau_1}} \\
\end{array}
\right).
\eea
We note that $\Gamma^1_{ab}$ is exact while various components of $\Gamma^2_{ab}$, $\Gamma^3_{ab}$ and $\Gamma^4_{ab}$ may have corrections of ${\cal O}({\cal V}^{-1})$. This shows the relevance of using the real field basis $\Phi^a =\{{\cal V}, t^2, t^3, t^4\}$ which is compatible with the need of creating the mass hierarchy among the overall volume modulus ${\cal V}$ and the remaining moduli which turn up as suitable candidates for assisted inflation. Further, for a three-field model with $\{{\cal V}, t^2, t^3\}$ without the fourth modulus $t^4$, one has $\tau_1 \to 2 t^2 t^3$ and ${\cal V} \to 2 t^1 t^2 t^3$, and subsequently one finds that the only non-vanishing affine connections are
\bea
& & \Gamma^1_{11} = - \frac{1}{\cal V}, \qquad \Gamma^2_{22} = - \frac{1}{t^2}, \qquad \Gamma^3_{33} = -\frac{1}{t^3},
\eea
which matches with the toroidal-like CY model presented in \cite{Leontaris:2025hly,Leontaris:2026sqh}.


\bibliographystyle{JHEP}
\bibliography{reference}


\end{document}